\documentclass{amsart}[10pt]

\usepackage{amscd,amssymb,latexsym,url,verbatim,graphicx,color}
\usepackage{tikz,tikz-cd}
\usetikzlibrary{decorations.pathmorphing}

\usepackage{cases,amsmath}

\usepackage{txfonts}

\usepackage{tikz,tikz-cd}

\usepackage[dvips]{epsfig}

\title
{Topology of the immediate snapshot complexes}

\author{Dmitry N. Kozlov}

\address{Department of Mathematics, University of Bremen, 28334
  Bremen, Federal Republic of Germany}

\email{dfk@math.uni-bremen.de}

\newtheorem{theorem}{Theorem}[section]
\newtheorem{df}[theorem]{Definition}
\newtheorem{thm}[theorem]{Theorem} 
\newtheorem{prop}[theorem]{Proposition}
\newtheorem{lm}[theorem]{Lemma}
\newtheorem{crl}[theorem]{Corollary}

\newtheorem{rem}[theorem]{Remark}
\newtheorem{dcc}[theorem]{Distributed Computing Context}
\newtheorem{conj}[theorem]{Conjecture} \newcommand{\nin}{\noindent}
\newcommand{\pr}{\nin{\bf Proof.} }

\newcommand{\act}{\text{\rm act}\,}

\newcommand{\ck}{{\mathcal K}}

\newcommand{\cn}{{\mathcal N}}

\newcommand{\da}{\Delta}
\newcommand{\dar}{\downarrow}

\newcommand{\es}{\emptyset}

\newcommand{\hra}{\hookrightarrow}
\newcommand{\inte}{\text{\rm int}\,}

\newcommand{\last}{\text{\rm last}\,}

\newcommand{\mqed}{\quad\Box}

\newcommand{\pass}{\text{\rm pass}\,}
\newcommand{\pnt}{round counter }

\newcommand{\ra}{\rightarrow}

\newcommand{\st}{\text{\rm{st}}}
\newcommand{\sm}{\setminus}
\newcommand{\smax}{\text{\rm smax}\,}

\newcommand{\supp}{\text{\rm supp}\,}

\newcommand{\tr}{{\bar r}}
\newcommand{\trc}{{\text{\rm Tr}}}

\newcommand{\wti}{\widetilde}
\newcommand{\zz}{{\mathbb Z}}
\newcommand{\ab}{\allowbreak}

\numberwithin{equation}{section}
\numberwithin{figure}{section}
\numberwithin{table}{section}

\begin{document}

\begin{abstract}
The immediate snapshot complexes were introduced as combinatorial
models for the protocol complexes in the context of theoretical
distributed computing. In the previous work we have developed a~formal
language of witness structures in order to define and to analyze
these complexes.

In this paper, we study topology of immediate snapshot complexes. It
is known that these complexes are always pure and that they are
pseudomanifolds. Here we prove two further independent topological
properties. First, we show that immediate snapshot complexes are
collapsible. Second, we show that these complexes are homeomorphic to
closed balls. Specifically, given any immediate snapshot complex
$P(\tr)$, we show that there exists a~homeomorphism
$\varphi:\da^{|\supp\tr|-1}\ra P(\tr)$, such that $\varphi(\sigma)$ is
a subcomplex of $P(\tr)$, whenever $\sigma$ is a simplex in the
simplicial complex $\da^{|\supp\tr|-1}$.
\end{abstract}

\maketitle

\section{Witness structures and immediate snapshot protocol complexes}

\subsection{Modeling protocol complexes for the immediate snapshot 
read/write distributed protocols} $\,$

\nin A crucial ingredient in the topological approach to theoretical
distributed computing, see~Herlihy et al, \cite{HKR}, is associating
a~simplicial complex, called the {\it protocol complex}, to every
distributed protocol, once the computational model is fixed. In this
paper, we study topology of standard full-information protocol
complexes in one of the central models of computation.

Let us fix the computational model to be the immediate snapshot
read/write model, which was originally introduced by Borowsky and
Gafni in~\cite{BG}. Roughly, this means that the processes can write
their values to the assigned memory registers, and they can read the
entire memory in one atomic step (snapshot read). The execution of the
protocol must have a layer structure, where in each layer a group of
processes becomes active, the processes in this group atomically write
their values to the memory, after this they atomically read the entire
memory.  Importantly, there are no further restrictions on how these
layers get activated during the protocol execution.

In our previous work, \cite{k1}, we introduced combinatorial models
for the protocol complexes for the standard protocols in that chosen
computational model, called {\it immediate snapshot complexes}. For
this, we needed to define new combinatorial structures, called {\it
  witness structures}, and study their structure theory, including
various operations, such as {\it ghosting}. We have proved that the
immediate snapshot complexes provide the correct model for these
protocol complexes, and started to study their topology. 

The standard protocols are naturally enumerated by finite sequences of
nonnegative integers, which we called {\it round counters}, denoted
$\tr$. Accordingly, the immediate snapshot complexes themselves were
denoted $P(\tr)$. In \cite{k1} it was proved that the complexes
$P(\tr)$ are always pseudomanifolds with boundary, and the
combinatorics of the boundary subcomplex was described.

In this paper, we improve our understanding of topology of $P(\tr)$
significantly. We refine the notion of canonical subcomplex
decomposition of $P(\tr)$ from \cite{k1}, and give a~complete
combinatorial description of the incidence relations in this
stratification. This gives us a~good approach to understanding the
inner structure of $P(\tr)$. In particular, it is straightforward to
prove the contractibility of $P(\tr)$ by pairing the combinatorial
description of this incidence structure with the standard result in
combinatorial topology, called the {\it Nerve Lemma},
see~\cite{book}. As a~first topological property we show a~stronger
result: namely, that the complexes $P(\tr)$ are always
collapsible. The collapsing sequence is also explicitly described.

It takes much more effort to derive the second topological property
of $P(\tr)$, namely the fact that each such complex is homeomorphic to
a~closed ball of dimension $|\supp\tr|-1$. This is the content of the
Corollary~\ref{crl:main}, which is an immediate consequence of our
main Theorem~\ref{thm:main}. Specifically, we prove that, for every
$P(\tr)$, there exists a~homeomorphism $\varphi:\da^{|\supp\tr|-1}\ra
P(\tr)$, such that $\varphi(\sigma)$ is a subcomplex of $P(\tr)$,
whenever $\sigma$ is a simplex in the simplicial complex
$\da^{|\supp\tr|-1}$.

The work presented here is the rigorous workout of the second part of
the preprint \cite{kfull}. The detailed expansion of the first part
of~\cite{kfull} has already appeared in \cite{k1}, where we laid the
combinatorial groundwork for the topological results of this paper. We
spend the rest of this section reminding the notations of~\cite{k1}
and results proved there. Our presentation here is quite condensed and
the reader is referred to~\cite{k1} for further details. We remark
that topology of protocol complexes for related computational models
has been studied by many authors, see
e.g.,~\cite{Ha04,HKR,HS,subd,view}. Furthermore, we recommend Attiya
and Welch, \cite{AW}, for an~in-depth background on theoretical
distributed computing.

Fundamentally, this paper can be viewed a~stand-alone article, written
in a~rigorous mathematical fashion, making it possible, in principle,
to be read independently. However, we strongly recommend that the
reader consults \cite{k1}, before starting reading this
one. Furthermore, in order both to facilitate researchers who are
mainly interested in distributed computing, as well as topologists
interested in more distributed computing background, we shall comment
throughout the text, explaining the distributed computing intuition
behind the mathematical concepts.

\subsection{Round counters} $\,$

\nin To start with, we review some of the standard terminology which
we will use. We let $\zz_+$ denote the set of nonnegative integers
$\{0,1,2,\dots\}$.  For a natural number $n$ we shall use $[n]$ to
denote the set $\{0,\dots,n\}$, with a convention that $[-1]=\es$.
For a~finite subset $S\subset\zz_+$, such that $|S|\geq 2$, we let
$\smax S$ denote the {\it second} largest element, i.e., $\smax
S:=\max(S\setminus\{\max S\})$. Finally, for a set $S$ and an element
$a$, we set
\[\chi(a,S):=
\begin{cases}
1, & \text{ if } a\in S;\\
0, & \text {otherwise.} 
\end{cases}\]
\nin Furthermore, whenever $(X_i)_{i=1}^t$ is a~family of topological
spaces, we set $X_I:=\bigcap_{i\in I} X_i$.  Also, when no confusion
arises, we identify one-element sets with that element, and write,
e.g., $p$ instead of $\{p\}$.

Next, we proceed to the combinatorial enumeration of all standard
protocols, together with relation terminology. This is accomplished by
the introduction of the so-called {\it round counters}.

\begin{df}
Given a function $\tr:\zz_+\rightarrow\zz_+\cup\{\bot\}$, we consider
the set
\[\supp\tr:=\{i\in\zz_+\,|\,\tr(i)\neq\bot\}.\] 
This set is called the {\bf support set} of $\tr$.

\nin A {\bf round counter} is a function
$\tr:\zz_+\rightarrow\zz_+\cup\{\bot\}$ with a~finite support set.
\end{df}

Obviously, a~round counter can be thought of as an infinite sequence
$\tr=(\tr(0),\tr(1),\dots)$, where, for all $i\in\zz_+$, either
$\tr(i)$ is a nonnegative integer, or $\tr(i)=\bot$, such that only
finitely many entries of $\tr$ are nonnegative integers. We shall
frequently use a~short-hand notation $\tr=(r_0,\dots,r_n)$ to denote
the round counter given by
\[\tr(i)=\begin{cases}
r_i, & \text{ for } 0\leq i\leq n;\\
\bot,& \text{ for } i>n.
\end{cases}\]

\begin{df}
Given a round counter $\tr$, the number $\sum_{i\in\supp\tr}\tr(i)$ is
called the {\bf cardinality} of $\tr$, and is denoted $|\tr|$. The
sets
\[\act\tr:=\{i\in\supp\tr\,|\,\tr(i)\geq 1\}
\text{ and } \pass\tr:=\{i\in\supp\tr\,|\,\tr(i)=0\}\] 
are called the {\bf active} and the {\bf passive} sets of~$\tr$.
\end{df}

\begin{dcc}
Since we consider full-information protocols only, they can be
described by specifying the number of rounds each process executes the
write-read sequence. Mathematically, these protocols are indexed by
round counters. Given a~round counter $\tr$, the set $\supp\tr$
indexes the participating processes, and is required to be finite. The
symbol $\bot$ means that the process does not participate.
Accordingly, the set $\pass\tr$ indexes the passive processes, i.e.,
those, which formally take part in the execution, but which do not
actually perform any active steps, while the set $\act\tr$ indexes the
processes which execute at least one step.
\end{dcc}

The following special class of round counters is important for our
study.

\begin{df}
For an arbitrary pair of disjoint finite sets $A,B\subseteq\zz_+$
we define a~round counter $\chi_{A,B}$ given by
\[\chi_{A,B}(i):=\begin{cases}
1, & \text{ if } i\in A;\\
0, & \text{ if } i\in B.
\end{cases}\]

Furthermore, for an arbitrary round counter $\tr$, we set
$\chi(\tr):=\chi_{\act\tr,\pass\tr}$.
\end{df}

We note that $\supp\tr=\supp(\chi(\tr))$. In the paper we shall also
use the short-hand notation $\chi_A:=\chi_{A,\es}$.

We define two operations on the round counters. To start with, assume
$\tr$ is a~\pnt and we have a~subset $A\subseteq\zz_+$. We let
$\tr\sm A$ denote the \pnt defined by
\[(\tr\sm A)(i)=\begin{cases}
\tr(i), & \text{ if } i\notin A;\\
\bot,   & \text{ if } i\in A.
\end{cases}\]
We say that the \pnt $\tr\sm A$ is obtained from $\tr$ by the {\it
  deletion} of~$A$. Note that $\supp(\tr\sm A)=\supp(\tr)\setminus S$,
$\act(\tr\sm A)=\act(\tr)\setminus A$, and $\pass(\tr\sm
A)=\pass(\tr)\setminus A$. 

Furthermore, we have $\chi(\tr\sm
A)=\chi(\tr)\sm A$. Finally, we note for future reference that for
$A\subseteq C\cup D$ we have
\begin{equation}\label{eq:chi2}
\chi_{C,D}\sm A=\chi_{C\sm A,D\sm A}.
\end{equation} 

For the second operation, assume $\tr$ is a~\pnt and we have a~subset
$S\subseteq\act\tr$. We let $\tr\dar S$ denote the \pnt defined by
\[(\tr\dar S)(i)=\begin{cases}
\tr(i), & \text{ if } i\notin S;\\
\tr(i)-1,   & \text{ if } i\in S.
\end{cases}\]
We say that the \pnt $\tr\dar S$ is obtained from $\tr$ by the {\it
  execution} of~$S$. Note that $\supp(\tr\dar S)=\supp\tr$,
$\act(\tr\dar S)=\{i\in\act\tr\,|\, i\notin S, \textrm{ or }\tr(i)\geq
2 \}$, and $\pass(\tr\dar S)=\pass(\tr)\cup\{i\in S\,|\,\tr(i)=1\}$.
However, in general we have $\chi(\tr)\dar S\neq\chi(\tr\dar S)$.

\begin{dcc} 
The replacement of $\tr$ with $\tr\sm A$ yields a new protocol, where
all processes from $A$ have been banned from participation. The
replacement of $\tr$ with $\tr\dar S$ corresponds to letting processes
from $S$ execute one round, and then running the remaining protocol
with new inputs.
\end{dcc}

For an arbitrary round pointer $\tr$ and sets $S\subseteq\act\tr$,
$A\subseteq\supp\tr$ we set
\begin{equation}\label{eq:rsa1}
\tr_{S,A}:=(\tr\dar S)\sm A=(\tr\sm A)\dar(S\sm A).
\end{equation}
In the special case, when $A\cap S=\es$, the identity \eqref{eq:rsa1}
specializes to
\begin{equation}\label{eq:rsa2}
\tr_{S,A}:=(\tr\dar S)\sm A=(\tr\sm A)\dar S.
\end{equation}
When $A=\es$, we shall frequently use the short-hand notation $\tr_S$
instead of $\tr_{S,A}$, in other words, $\tr_S=\tr\dar S$. 
Again, for future reference, we note that for $S\subseteq C$, we have
\begin{equation}\label{eq:chi3}
\chi_{C,D}\dar S=\chi_{C\sm S,D\cup S}.
\end{equation}

\subsection{Witness structures and the ghosting operation} $\,$

\nin Next, we describe the basic terminology which we will need to
define the immediate snapshot complexes.

\begin{df}\label{df:ws}
A~{\bf witness prestructure} is a~finite sequence of pairs of finite
subsets of $\zz_+$, denoted $\sigma=((W_0,G_0),\dots,(W_t,G_t))$, with
$t\geq 0$, satisfying the following conditions:
\begin{enumerate}
\item[(P1)] $W_i,G_i\subseteq W_0$, for all $i=1,\dots,t$;
\item[(P2)] $G_i\cap G_j=\emptyset$, for all $i,j\in[t]$, $i<j$;
\item[(P3)] $G_i\cap W_j=\emptyset$, for all $i,j\in[t]$, $i\leq j$.
\end{enumerate}

\noindent
A witness prestructure is called {\bf stable} if in addition the
following condition is satisfied:
\begin{enumerate}
\item[(S)] if $t\geq 1$, then $W_t\neq\emptyset$.
\end{enumerate}

\noindent
A {\bf witness structure} is a witness prestructure satisfying the
following strengthening of condition (S):
\begin{enumerate}
\item[(W)] the subsets $W_1,\dots,W_t$ are all nonempty.
\end{enumerate}
\end{df}

\begin{df}
We define the following data associated to an arbitrary witness
prestructure $\sigma=((W_0,G_0),\dots,(W_t,G_t))$:
\begin{itemize}
\item the set $W_0\cup G_0$ is called the {\bf support} of $\sigma$ and is
denoted by $\supp\sigma$;
\item the {\bf ghost set} of $\sigma$ is the set
  $G(\sigma):=G_0\cup\dots\cup G_t$;
\item the {\bf active set} of $\sigma$ is the complement of the ghost set
\[A(\sigma):=\supp(\sigma)\setminus G(\sigma)=W_0\setminus(G_1\cup\dots\cup G_t);\]
\item the {\bf dimension} of $\sigma$ is
  \[\dim\sigma:=|A(\sigma)|-1=|W_0|-|G_1|-\dots-|G_t|-1.\]
\end{itemize}
\end{df}
\noindent
For brevity of some formulas, we set $W_{-1}:=W_0\cup G_0=\supp\sigma$.

\begin{df}\label{df:trc}
For a~prestructure $\sigma$ and an arbitrary $p\in\supp\sigma$, we set
\[\trc(p,\sigma):=\{0\leq i\leq t\,|\,p\in W_i\cup G_i\},\] 
and call it the {\bf trace} of~$p$. Furthermore, for all
$p\in\supp\sigma$, we set 
\[\last(p,\sigma):=\max\{-1\leq i\leq t\,|\,p\in W_i\}.\]
\end{df}

\nin When the choice of $\sigma$ is unambiguous, we shall simply write
$\trc(p)$ and $\last(p)$. The following definition provides an
alternative approach to witness structures using traces.

\begin{df}\label{df:trws}
A~{\bf witness prestructure} is a~pair of finite subsets
$A,G\subseteq\zz_+$ together with a~family $\{\trc(p)\}_{p\in A\cup
  G}$ of finite subsets of $\zz_+$, satisfying the following
condition:
\begin{itemize}
\item[(T)] $0\in\trc(p)$, for all $p\in A\cup G$.
\end{itemize}

\noindent
A witness prestructure is called {\bf stable} if it satisfies
an~additional condition:
\begin{itemize}
\item[(TS)] if $A=\emptyset$, then $\trc(p)=\{0\}$, for all $p\in G$,
else 
\[\max_{p\in A}\last(p)\geq\max\bigcup_{p\in G}\trc(p).\]
\end{itemize}

Set $t:=\max_{p\in A}\last(p)$. A~stable witness prestructure is
called {\bf witness structure} if the following strengthening of
Condition (TS) is satisfied:
\begin{enumerate}
\item[(TW)] {\it for all $1\leq k\leq t$ either there exists $p\in A$
  such that $k\in\trc(p)$, or there exists $p\in G$ such that
  $k\in\trc(p)\sm\max\trc(p)$.}
\end{enumerate}
\end{df}

We shall call the form of the presentation of the witness prestructure
as a~triple $(A,G,\{\trc(p)\}_{p\in A\cup G})$ its {\it trace form}.

\begin{dcc}
The witness structure is a mathematical object modelling the
information which the processes have during the execution of the
full-information protocol. Let us explain the distributed computing
intuition behind this notation.

The set $\supp\sigma$ indexes all processes which are participating in
the protocol. The processes indexed by the set $W_0$ are of two
different types.  Those, whose view is included in $\sigma$, and
those, who have only been passively witnessed by others. The processes
of the first type are indexed by the set $A(\sigma)$, the other ones
are indexed by the union $G_1\cup\dots\cup G_t$. The set $G_0$ indexes
those processes from $\supp\sigma$ which have not be witnessed by
anybody in this particular execution.

The fact, that $p\in W_k$ is to be interpreted as ``the active
participation of process $p$ in round $k$ has been witnessed''. This
can happen in two ways, either $p$ itself is active in this execution,
or $p$ is being passively witnessed and this is not the last occurence
of $p$. The fact that $p\in G_k$ means that process $p$ has been
passively witnessed and this is the last occurence of $p$. 

We refer the reader to \cite[Section 6]{k1}, where connection between
witness structures and witness posets is explained.
\end{dcc}

Next, we proceed to describe various operations in witness structures
and prestructures. To start with, any stable witness prestructure can
be turned into a witness structure, which is called its {\it canonical
  form}.

\begin{df}\label{df:cform}
Assume $\sigma=((W_0,G_0),\dots,(W_t,G_t))$ is an arbitrary stable
witness prestructure. Set $q:=|\{1\leq i\leq t\,|\,
W_i\neq\emptyset\}|$.  Pick $0=i_0<i_1<\dots<i_q=t$, such that
$\{i_1,\dots,i_q\}=\{1\leq i\leq t\,|\,W_i\neq\emptyset\}$.  We define
the witness structure $C(\sigma)=((W_0,G_0),\ab(\wti W_1, \wti
G_1),\allowbreak \dots,\ab(\wti W_q,\wti G_q))$, which is called the {\bf
  canonical form} of~$\sigma$, by setting
\begin{equation}\label{eq:canon}
\wti W_k:=W_{i_k},\quad 
\wti G_k:=G_{i_{k-1}+1}\cup\dots\cup G_{i_k},\text{ for all } k=1,\dots,q,
\end{equation}
\end{df}

Furthermore, any witness prestructure can be made stable using the
following operation.

\begin{df} \label{df:st}
Let $\sigma=((W_0,G_0),\dots,(W_t,G_t))$ be a~witness prestructure,
set
\[q:=\max\{0\leq i\leq t\,|\,W_i\not\subseteq G(\sigma)\}.\]
The {\bf stabilization} of $\sigma$ is the witness prestructure
$\st(\sigma)$ whose trace form is $(A(\sigma),\ab G(\sigma), \ab
\{\trc(p)|_{[q]}\}_{p\in\supp\sigma})$.

More generally, assume $S\subseteq A(\sigma)$, and set
\[q:=\max\{0\leq i\leq t\,|\,W_i\not\subseteq S\cup G(\sigma)\}.\]
The {\bf stabilization} of $\sigma$ {\bf modulo} $S$ is the witness
prestructure $\st_S(\sigma)$ whose trace form is $(A(\sigma)\setminus
S,\ab G(\sigma)\cup S, \{\trc(p)|_{[q]}\}_{p\in\supp\sigma})$.
\end{df}

Combining stabilization modulo a~set with taking the canonical form
yields a~new operation, called {\it ghosting}, which will be of utter
importance for the combinatorial description of the incidence
structure in the immediate snapshot complexes.

\begin{df}\label{df:go}
For an arbitrary witness structure $\sigma$, and an~arbitrary
$S\subseteq A(\sigma)$, we define
$\Gamma_S(\sigma):=C(\st_S(\sigma))$. We say that $\Gamma_S(\sigma)$
is obtained from $\sigma$ {\bf by ghosting~$S$}.
\end{df}

\begin{dcc}\label{dcc:4}
The operation of ghosting the set of processes $S$ corresponds to
excluding their views from the knowledge that the witness structure
encodes. Clearly, the occurences of processes from $S$ will not vanish
from the witness structure $\Gamma_S(\sigma)$ altogether, but these
processes will cease being active, and whatever we will see of them
will just be the residual information passively witnessed by other
processes.
\end{dcc}

The main property of ghosting which one needs for proving the
well-definedness of the immediate snapshot complexes is that it
behaves well with respect to iterations.

\begin{prop} 
\label{prop:gg}
Assume $\sigma$ is a witness structure, and $S,T\subseteq A(\sigma)$, 
such that $S\cap T=\emptyset$. Then we have
$\Gamma_T(\Gamma_S(\sigma))=\Gamma_{S\cup T}(\sigma)$, expressed 
functorially we have $\Gamma_T\circ\Gamma_S=\Gamma_{S\cup T}$.
\end{prop}

\subsection{Immediate snapshot complexes} $\,$

\nin We have now introduced sufficient terminology in order to
describe our main objects of study.

\begin{df}\label{df:ptr}
Assume $\tr$ is a~round counter. We define an abstract simplicial
complex $P(\tr)$, called the {\bf immediate snapshot complex}
associated to the round counter~$\tr$, as follows. The vertices of
$P(\tr)$ are indexed by all witness structures $\sigma=(\{p\},\ab
G,\ab \{\trc(q)\}_{q\in\{p\}\cup G})$, satisfying these three
conditions:
\begin{enumerate}
\item $\{p\}\cup G=\supp\tr$;
\item $|\trc(p)|=r(p)+1$;
\item $|\trc(q)|\leq r(q)+1$, for all $q\in G$.
\end{enumerate}
\nin We say that such a vertex has {\bf color} $p$. In general, the
simplices of $P(\tr)$ are indexed by all witness structures
$\sigma=(A,\ab G,\ab \{\trc(q)\}_{q\in A\cup G})$, satisfying:
\begin{enumerate}
\item $A\cup G=\supp\tr$;
\item $|\trc(q)|=r(q)+1$, for all $q\in A$;
\item $|\trc(q)|\leq r(q)+1$, for all $q\in G$.
\end{enumerate}

\nin The empty witness structure $((\es,\supp\tr))$ indexes the empty
simplex of $P(\tr)$. When convenient, we identify the simplices of
$P(\tr)$ with the witness structures which index them.  

Let $\sigma$ be a~non-empty witness structure satisfying the
conditions above. The set of vertices $V(\sigma)$ of the simplex $\sigma$
is given by $\{\Gamma_{A(\sigma)\setminus\{p\}}(\sigma)\,|\,p\in A\}$.
\end{df}

Taking boundaries of simplices in $P(\tr)$ corresponds to ghosting of
the witness structures. This is only natural taking into account the
intuition from the distributed computing context~\ref{dcc:4}.

\begin{prop}
\label{prop:b}
Assume $\tr$ is the round counter, and assume $\sigma$ and $\tau$ are
simplices of $P(\tr)$. Then $\tau\subseteq\sigma$ if and only if there
exists $S\subseteq A(\sigma)$, such that $\tau=\Gamma_S(\sigma)$.
\end{prop}

The first property of the simplicial complexes $P(\tr)$, which is
quite easy to see, is that these complexes are pure of dimension
$|\supp\tr|-1$. Furthermore, zero values in the round counter have
a~simple topological interpretation.

\begin{prop}\label{prop:p4}{\rm (\cite[Proposition 4.4]{k1}).}
Assume $\tr=(r(0),\dots,r(n))$ and $\tr(n)=0$. Let $\bar q$ denote the
truncated round counter $(r(0),\dots,r(n-1))$. Consider a~cone over
$P(\bar q)$, which we denote $P(\bar q)*\{a\}$, where $a$ is the apex
of the cone. Then we have
\begin{equation}\label{eq:ptr} 
P(\tr)\simeq P(\bar q)*\{a\},
\end{equation} 
where $\simeq$ denotes the simplicial isomorphism.
\end{prop}

For brevity, we set $P_n:=P(\underbrace{1,\dots,1}_{n+1})$. It turns
out that the standard chromatic subdivision of an~$n$-simplex,
see~\cite{subd}, is a~special case of the immediate snapshot complex.

\begin{prop} {\rm (\cite[Proposition 4.10]{k1}).}
The immediate snapshot complex $P_n$ and the standard chromatic
subdivision of an $n$-simplex $\chi(\da^n)$ are isomorphic as
simplicial complexes. Explicitly, the isomorphism can be given by
\begin{equation}\label{eq:bc}
\Phi:((B_1,\dots,B_t)(C_1,\dots,C_t))\mapsto
\begin{array}{|c|c|c|c|c|}
\hline
W_0 & C_1        & C_2        & \dots & C_t \\ \hline
[n]\sm W_0 & B_1\sm C_1 & B_2\sm C_2 & \dots & B_t\sm C_t \\ 
\hline
\end{array},
\end{equation}
where $W_0=B_1\cup\dots\cup B_t$.
\end{prop}

Recall the following property of pure simplicial complexes,
strengthening the usual connectivity.

\begin{df}
Let $K$ be a pure simplicial complex of dimension $n$. Two
$n$-simplices of $K$ are said to be {\bf strongly connected} if there
is a~sequence of $n$-simplices so that each pair of consecutive
simplices has a~common $(n-1)$-dimensional face. The complex $K$ is
said to be {\bf strongly connected} if any two $n$-simplices of $K$
are strongly connected.
\end{df}

Clearly, being strongly connected is an equivalence relation on the
set of all $n$-simplices.

\begin{prop}\label{prop:strc} {\rm (\cite[Proposition 5.6]{k1}).}
For an arbitrary round counter $\tr$, the simplicial complex $P(\tr)$
is strongly connected.
\end{prop}

The next definition describes a~weak simplicial analog of being
a~manifold.

\begin{df}
We say that a~strongly connected pure simplicial complex $K$ is a~{\bf
  pseudomanifold} if each $(n-1)$-simplex of $K$ is a~face of
precisely one or two $n$-simplices of $K$. The $(n-1)$-simplices of
$K$ which are faces of precisely one $n$-simplex of $K$ form
a~simplicial subcomplex of $K$, called the {\bf boundary} of $K$, and
denoted $\partial K$.
\end{df}

It was shown in \cite{k1}, that immediate snapshot complexes are
always pseudomanifolds with boundary.

\begin{thm}\label{prop:pseudo} {\rm (\cite[Proposition 5.9]{k1}).}
For an arbitrary round counter $\tr$, the simplicial complex $P(\tr)$
is a~pseudomanifold, where $\partial P(\tr)$ is the subcomplex
consisting of all simplices $\sigma=((W_0,G_0),\ab\dots,(W_t,G_t))$,
satisfying $G_0\neq\es$.
\end{thm}

\section{A canonical decomposition of the immediate snapshot complexes}

\subsection{Definition and examples}
$\,$

\nin The canonical decomposition of the immediate snapshot complexes has
been introduced in \cite{k1}. In order to better understand the
topology of these complexes, we need to generalize that definition and
look at finer strata.

\begin{df}
Assume $\tr$ is a~round counter.
\begin{itemize}
\item For every subset $S\subseteq\act\tr$, let $Z_S$ denote the set
  of all simplices $\sigma=((W_0,G_0),\ab\dots,\ab(W_t,G_t))$, such
  that $S\subseteq G_1$.
\item For every pair of subsets $A\subseteq S\subseteq\act\tr$, let
  $Y_{S,A}$ denote the set of all simplices
  $\sigma=((W_0,G_0),\dots,(W_t,G_t))$, such that $R_1=S$ and
  $A\subseteq G_1$. Furthermore, set $X_{S,A}:=Y_{S,A}\cup Z_S$
\end{itemize} 
\end{df}

We shall also use the following short-hand notation:
$X_S:=X_{S,\emptyset}$. This case has been considered in \cite{k1},
where it was shown that $X_S$ is a~simplicial subcomplex of $P(\tr)$
for an~arbitrary~$S$.

\begin{dcc}
The subcomplexes $X_S$ correspond to the subset of executions which
start with the processes indexed by the set $S$ executing
simultaneously.  This explains, from the point of view of distributed
computing, why the protocol complex decomposes into these strata.
\end{dcc}

On the other extreme, clearly $Z_S=X_{S,S}$ for all $S$.  When
$A\not\subseteq S$, we shall use the convention $Y_{S,A}=\es$.  Note,
that in general the sets $Y_{S,A}$ need not be closed under taking
boundary.

\begin{prop}
The sets $X_{S,A}$ are closed under taking boundary, hence form
simplicial subcomplexes of $P(\tr)$.
\end{prop}
\pr Let $\sigma=((W_0,G_0),\dots,(W_t,G_t))$ be a~simplex in
$X_{S,A}$, and assume $\tau\subset\sigma$. By Proposition~\ref{prop:b}
there exists $T\subseteq A(\sigma)$, such that
$\tau=\Gamma_T(\sigma)$. By Proposition~\ref{prop:gg} it is enough to
consider the case $|T|=1$, so assume $T=\{p\}$, and let $\tau=((\wti
W_0,\wti G_0),\dots,(\wti W_{\tilde t},\wti G_{\tilde t}))$.

By definition of $X_{S,A}$ we have either $\sigma\in Z_S$ or
$\sigma\in Y_{S,A}$.  Consider first the case $\sigma\in Z_S$, so
$S\subseteq G_1$. Since $\wti G_1\supseteq G_1$, we have $\tau\in
Z_S$.

Now, assume $\sigma\in Y_{S,A}$. This means $W_1\cup G_1=S$ and
$A\subseteq G_1$. Again $\wti G_1\supseteq G_1$ implies
$A\subseteq\wti G_1$.  \qed

\vspace{5pt}

\noindent
In particular, $X_S$ and $Z_S$ are simplicial subcomplexes of
$P(\tr)$, for all $S$. When we are dealing with several round
counters, in order to avoid confusion, we shall add $\tr$ to the
notations, and write $X_{S,A}(\tr)$, $X_S(\tr)$, $Y_{S,A}(\tr)$,
$Z_S(\tr)$. We shall also let $\alpha_{S,A}(\tr)$ denote the inclusion
map 
\[\alpha_{S,A}(\tr):X_{S,A}(\tr)\hookrightarrow P(\tr).\]


\subsection{The strata of the canonical decomposition as immediate snapshot complexes}
$\,$

\nin The first important property of the simplicial complexes
$X_{S,A}$ is that they themselves can be interpreted as immediate
snapshot complexes. Here, and in the rest of the paper, we shall use
$\rightsquigarrow$ to denote simplicial isomorphisms.

\begin{prop}\label{prop:strata}
Assume $A\subseteq S\subseteq\act\tr$, then there exists a~simplicial
isomorphism
\[\gamma_{S,A}(\tr):X_{S,A}(\tr)\rightsquigarrow P(\bar r_{S,A}).\] 
\end{prop}
\pr We start by considering the case $A=\es$. Pick an~arbitrary
simplex $\sigma=((W_0,G_0),\allowbreak \dots,(W_t,G_t))$ belonging to $X_S$.  
By the construction of $X_S$, we either have $W_1\cup G_1=S$, or
$S\subseteq G_1$. If $W_1\cup G_1=S$, then set
\[\gamma_S(\sigma):=\begin{array}{|c|c|c|c|}
\hline
W_0\sm G_1  & W_2 & \dots & W_t \\ \hline
G_0\cup G_1 & G_2 & \dots & G_t \\ 
\hline
\end{array},\]
else $S\subseteq G_1$, in which case we set
\[\gamma_S(\sigma):=\begin{array}{|c|c|c|c|c|}
\hline
W_0\sm  S & W_1      & W_2 & \dots & W_t \\ \hline
G_0\cup S & G_1\sm S & G_2 & \dots & G_t \\ 
\hline
\end{array}.\]

Reversely, assume $\tau=((V_0,H_0),\dots,(V_t,H_t))$ is a~simplex of
$P(\bar r_S)$. Note, that in any case, we have $S\subseteq V_0\cup
H_0$. If $V_0\cap S\neq\es$, we set
\[\rho_S(\tau):=\begin{array}{|c|c|c|c|c|}
\hline
V_0\cup(H_0\cap S) & V_0\cap S & V_1 & \dots & V_t \\ \hline
H_0\sm(H_0\cap S)  & H_0\cap S & H_1 & \dots & H_t \\ 
\hline
\end{array},\]
else $S\subseteq H_0$, and we set
\[\rho_S(\tau):=\begin{array}{|c|c|c|c|c|}
\hline
V_0\cup S & V_1       & V_2 & \dots & V_t \\ \hline
H_0\sm  S & H_1\cup S & H_2 & \dots & H_t \\ 
\hline
\end{array}.\]

It is immediate that $\gamma_S$ and $\rho_S$ preserve supports,
$A(-)$, $G(-)$, and hence also the dimension. Furthermore, we can see
what happens with the cardinalities of the traces. For all elements
$p$ which do not belong to~$S$, the cardinalities of their traces are
preserved. For all elements in $S$, the map $\gamma_S$ decreases the
cardinality of the trace, whereas, the map $\rho_S$ increases it. It
follows that $\gamma_S$ and $\rho_S$ are well-defined as
dimension-preserving maps between sets of simplices.

To see that $\gamma_S$ preserves boundaries, pick a~top-dimensional
simplex $\sigma=(W_0,S,\allowbreak W_1,\dots,W_t)$ in $X_S$ and ghost the
set~$T$. Assume first $S\not\subseteq T$. In this case not all
elements in $S$ are ghosted. Assume now that $S\subseteq T$. This
implies that $\gamma_S$ is well-defined as a~simplicial map.  Finally,
a~direct verification shows that the maps $\gamma_S$ and $\rho_S$ are
inverses of each other, hence they are simplicial isomorphisms.

Let us now consider the case when $A$ is arbitrary. The simplicial
complex $X_{S,A}$ is a~subcomplex of $X_S$ consisting of all simplices
$\sigma$ satisfying the additional condition $A\subseteq G_1$. The
image $\gamma_S(X_{S,A})$ consists of all
$\tau=((V_0,H_0),\dots,(V_t,H_t))$ in $P(\bar r_{S,A})$ satisfying
$A\subseteq H_0$. The map $\Xi:\gamma_S(X_{S,A})\rightarrow P(\bar
r_{S,A})$, taking $\tau$ to $((V_0,H_0\setminus
A),(V_1,H_1),\dots,(V_t,H_t))$, is obviously a~simplicial isomorphism,
hence the composition
$\gamma_{S,A}=\Xi\circ\gamma_S:X_{S,A}\rightarrow P(\bar r_{S,A})$ is
a~simplicial isomorphism as well.  \qed

\vskip5pt

\nin Note that, in particular,
\[\gamma_{A,A}(\sigma)=\begin{array}{|c|c|c|c|c|}
\hline
W_0\sm A & W_1      & W_2 & \dots & W_t \\ \hline
G_0      & G_1\sm A & G_2 & \dots & G_t \\ 
\hline
\end{array}.\]
\nin The statement of Proposition~\ref{prop:strata} for the example
$\tr=(2,1,1)$, is shown on Figure~\ref{fig:f211b}.

The next proposition is a~first of several results, which claim
commutativity of a~certain diagram. All these results have alternative
intuitive meaning. For example, the commutativity of the
diagram~\eqref{cd:xs} can be interpreted as saying that the relation of
the stratum $X_{S\cup A,A}(\tr)$ to $X_{A,A}(\tr)$ is the same as the
relation of the stratum $X_S(\tr\sm A)$ to $P(\tr\sm A)$.

\begin{prop}\label{prop:cxs}
Assume $\tr$ is an arbitrary round counter, and $S,A\subset\act\tr$,
such that $S\cap A=\es$, then the following diagram commutes
\begin{equation}\label{cd:xs}
\begin{tikzcd}[column sep=large]
X_{A,A}(\tr)\arrow[hookleftarrow]{r}{i}
\arrow[squiggly]{d}[swap]{\gamma_{A,A}(\tr)}
&X_{S\cup A,A}(\tr)\arrow[squiggly]{rd}{\gamma_{S\cup A,A}(\tr)} \\
P(\tr\sm A)\arrow[hookleftarrow]{r}{\alpha_S(\tr\sm A)}
& X_S(\tr\sm A)\arrow[squiggly]{r}{\gamma_S(\tr\sm A)}
& P(\tr_{S,A}),
\end{tikzcd}
\end{equation}
where $i$ denotes the strata inclusion map.
\end{prop}
\pr To start with, note that $\tr_{S,A}=(\tr\dar S)\sm
A=(\tr\dar(S\cup A))\sm A$, so the diagram \eqref{cd:xs} is
well-defined. To see that it is commutative, pick an arbitrary
$\sigma=((W_0,G_0),\dots,(W_t,G_t))$. We know that either $A\subseteq
G_1$ and $W_1\cup G_1=S\cup A$, or $A\cup S\subseteq G_1$. On one
hand, we have
\[(\gamma_{A,A}(\tr)\circ i)(\sigma)=\begin{array}{|c|c|c|c|c|}
\hline
W_0\sm A & W_1      & W_2 & \dots & W_t \\ \hline
G_0      & G_1\sm A & G_2 & \dots & G_t \\ 
\hline
\end{array}.\]
On the other hand, we have
\[\gamma_{S\cup A,A}(\tr)(\sigma)=
\begin{cases}
\begin{array}{|c|c|c|c|}
\hline
W_0\sm G_1       &  W_2 & \dots & W_t \\ \hline
G_0\cup G_1\sm A &  G_2 & \dots & G_t \\ 
\hline
\end{array}, 
\textrm{ if } A\subseteq G_1, W_1\cup G_1=S\cup A;\\[0.6cm]
\begin{array}{|c|c|c|c|c|}
\hline
W_0\sm(S\cup A) & W_1             & W_2 & \dots & W_t \\ \hline
G_0\cup S       & G_1\sm(S\cup A) & G_2 & \dots & G_t \\ 
\hline
\end{array},
\textrm{ if } A\cup S \subseteq G_1.
\end{cases}\]
Applying $\gamma_S(\tr\sm A)^{-1}$ we can verify that
$\gamma_{A,A}(\tr)\circ i=\alpha_S(\tr\sm A)\circ\gamma_S(\tr\sm
A)^{-1}\circ\gamma_{S\cup A,A}(\tr)$.  \qed

\begin{crl}\label{crl:xaa}
 For any $A\subseteq\act\tr$, we have
\begin{equation}\label{eq:xaa}
X_{A,A}(\tr)=\bigcup_{\es\neq S\subseteq\act\tr\sm A}X_{S\cup A,A}(\tr)=
\bigcup_{A\subset T\subseteq\act\tr}X_{T,A}(\tr).
\end{equation}
\end{crl}
\pr Since $P(\tr\sm A)=\bigcup_{\es\neq S\subseteq\act\tr\sm
  A}X_S(\tr\sm A)$, the equation~\eqref{eq:xaa} is an immediate
consequence of the commutativity of the diagram~\eqref{cd:xs}. \qed

\subsection{The incidence structure of the canonical decomposition} $\,$

\nin Clearly, $P(\tr)=\cup_S X_S(\tr)$. We describe here the complete
combinatorics of intersecting these strata.

\begin{prop} \label{pr:inc1}
For all pairs of subsets $A\subseteq S\subseteq\supp\tr$ and
$B\subseteq T\subseteq\supp\tr$ we have: $X_{S,A}\subseteq X_{T,B}$ if
and only if at least one of the following two conditions is satisfied:
\begin{itemize}
\item $S=T$ and $B\subseteq A$,
\item $T\subseteq A$.
\end{itemize} 
\end{prop}

\noindent
We remark that it can actually happen that both conditions in
Proposition~\ref{pr:inc1} are satisfied. This happens exactly when
$S=T=A$.

\vspace{5pt}

\noindent
{\bf Proof of Proposition~\ref{pr:inc1}.}  First we show that
$T\subseteq A$ implies $X_{S,A}\subseteq X_{T,B}$.  Take $\sigma\in
X_{S,A}$. If $\sigma\in Z_S$, then we have the following chain of
implications: $S\subseteq G_1\Rightarrow A\subseteq G_1\Rightarrow
T\subseteq G_1\Rightarrow\sigma\in Z_T$. If, on the other hand,
$\sigma\in Y_{S,A}$, we also have $A\subseteq G_1$, implying
$T\subseteq G_1$, hence $\sigma\in Z_T$.

Next we show that if $S=T$ and $B\subseteq A$, then $X_{S,A}\subseteq
X_{S,B}$.  Clearly, we just need to show that $Y_{S,A}\subseteq
X_{S,B}$.  Take $\sigma\in Y_{S,A}$, then we have the following chain
of implications:
\[\begin{cases}R_1=S\\A\subseteq G_1\end{cases}\Rightarrow
\begin{cases}R_1=T\\B\subseteq G_1\end{cases}
\Rightarrow\sigma\in Y_{T,B}.\]
This proves the {\it if} part of the proposition.

To prove the {\it only if} part, assume $X_{S,A}\subseteq X_{T,B}$. If
$S\neq A$, set
\[\tau:=\begin{array}{|c|c|c|c|c|}
\hline
\supp\tr & S\sm A & p_1 & \dots & p_t \\ \hline
\es      & A      & \es & \dots & \es \\ 
\hline
\end{array},\]
else $S=T$, and we set
\[\tau:=\begin{array}{|c|c|c|c|c|}
\hline
\supp\tr & p_1 & p_2 & \dots & p_t \\ \hline
\es      & S   & \es & \dots & \es \\ 
\hline
\end{array},\]
where in both cases $p_1,\dots,p_t$ is a sequence of elements from
$\supp\tr\sm A$, with each element $p$ occurring $\tr(p)$
times. Clearly, in the first case, $\tau\in Y_{S,A}$, and in the
second case $\tau\in Z_S$, hence $\tau\in X_{T,B}=Z_T\cup Y_{T,B}$.
This means that either $T\subseteq A$, or $S=T$ and $B\subseteq A$.
\qed

\begin{lm}\label{lm:yzint}
Assume $A\subseteq S\subseteq\supp\tr$ and
$B\subseteq T\subseteq\supp\tr$. We have
\begin{enumerate}
	\item [(1)] $Z_S\cap Z_T=Z_{S\cup T}$,
	\item [(2)] $Y_{S,A}\cap Z_T=Y_{S,A\cup T}$,
	\item [(3)] $Y_{S,A}\cap Y_{T,B}=\begin{cases}
	Y_{S,A\cup B}, & \textrm{ if } S=T,\\ 
	\es, & \textrm{ otherwise}.\end{cases}$
\end{enumerate}
\end{lm}
\pr To show (1), pick $\sigma\in Z_S\cap Z_T$. We have $S\subseteq
G_1$ and $T\subseteq G_1$, hence $S\cup T\subseteq G_1$, and so
$\sigma\in Z_{S\cup T}$.

To show (2), pick $\sigma\in Y_{S,A}\cap Z_T$. We have $R_1=S$,
$A\subseteq G_1$, and $T\subseteq G_1$. It follows that $R_1=S$ and
$A\cup T\subseteq G_1$, so $\sigma\in Y_{S,A\cup T}$.

Finally, to show (3), pick $\sigma\in Y_{S,A}\cap Y_{T,B}$. On one
hand, $\sigma\in Y_{S,A}$ means $R_1=S$ and $A\subseteq G_1$, on the
other hand, $\sigma\in Y_{T,B}$ means $R_1=T$ and $B\subseteq G_1$. We
conclude that $Y_{S,A}\cap Y_{T,B}=\es$ if $S\neq T$. Otherwise, we
have $R_1=S=T$ and $A\cup B\subseteq G_1$, so $\sigma\in Y_{S,A\cup
  B}$.  \qed

\begin{prop} \label{pr:inc2}
For all pairs of subsets $A\subseteq S\subseteq\supp\tr$ and
$B\subseteq T\subseteq\supp\tr$ we have the following formulae for the
intersection:
\begin{numcases}
{X_{S,A}\cap X_{T,B}=}
X_{S,A\cup B}, & if  $S=T$; \label{inc:1} \\
X_{T,S\cup B}, & if  $S\subset T$; \label{inc:2}\\
Z_{S\cup T}=X_{S\cup T,S\cup T}, & if  
$S\not\subseteq T$ and $T\not\subseteq S$. \label{inc:3}
\end{numcases}
\end{prop}
\pr In general, we have
\begin{multline} \label{eq:xx}
X_{S,A}\cap X_{T,B}=(Z_S\cap Z_T)\cup(Z_S\cap Y_{T,B})\cup(Y_{S,A}\cap Z_T)\cup
(Y_{S,A}\cap Y_{T,B})\\
=\begin{cases}
Z_{S\cup T}\cup Y_{T,S\cup B}\cup Y_{S,T\cup A}\cup Y_{S,A\cup B}, & \text{ if } S=T; \\
Z_{S\cup T}\cup Y_{T,S\cup B}\cup Y_{S,T\cup A}, & \text{ otherwise}.
\end{cases}
\end{multline}

Assume first that $S=T$. In this case $Y_{T,S\cup B}=Y_{S,T\cup
  A}=Z_S$, hence the equation \eqref{eq:xx} translates to $X_{S,A}\cap
X_{T,B}=Z_S\cup Y_{S,A\cup B}=X_{S,A\cup B}$.

Let us now consider the case $S\subset T$. We have $Y_{S,T\cup
  A}=\es$, hence \eqref{eq:xx} translates to $X_{S,A}\cap
X_{T,B}=Z_T\cup Y_{T,S\cup B}=X_{T,S\cup B}$.

Finally, assume $S\not\subseteq T$ and $T\not\subseteq S$. Then
$Y_{T,S\cup B}=Y_{S,T\cup A}=\es$, hence \eqref{eq:xx} says
$X_{S,A}\cap X_{T,B}=Z_{S\cup T}$.  \qed

For convenience we record the following special cases of
Proposition~\ref{pr:inc2}.

\begin{crl}\label{crl:6.6}
For $S\neq T$ we have
\[X_S\cap X_T=\begin{cases}
X_{T,S}, & \text{ if } S\subset T, \\
Z_{S\cup T}, & \text{ otherwise,}
\end{cases}\]
\begin{equation}
\label{eq:xz}
X_S\cap Z_T=Z_{S\cup T}.
\end{equation}
\end{crl}
\pr The first formula is a simple substitution of $A=B=\es$ in
\eqref{inc:2} and \eqref{inc:3}. To see \eqref{eq:xz}, substitute
$A=\es$, $B=T$ in \eqref{inc:2} to obtain
\[X_{S,\es}\cap X_{T,T}=\begin{cases}
X_{T,S\cup T}, & \text{ if } S\subset T \\
Z_{S\cup T}, & \text{ otherwise}
\end{cases}=\begin{cases}
Z_T, & \text{ if } S\subset T \\
Z_{S\cup T}, & \text{ otherwise}
\end{cases}=Z_{S\cup T}.\mqed
\]

\begin{figure}[hbt]

  \input{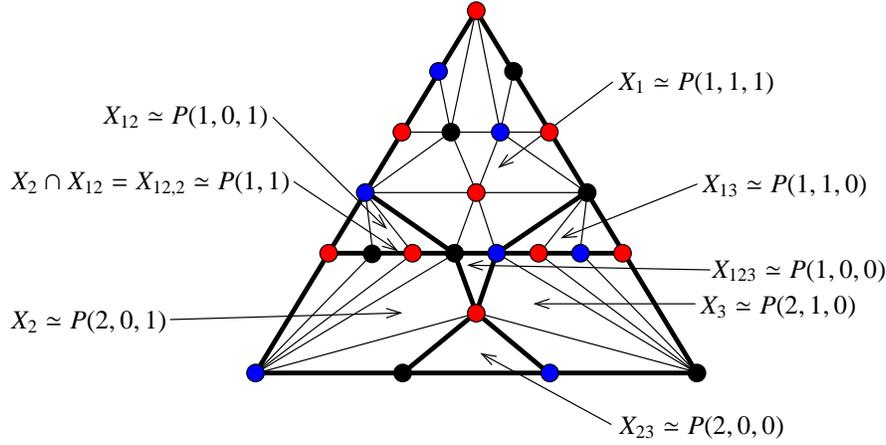}  

\caption{The canonical decomposition of the immediate snapshot complex $P(2,1,1)$.}
\label{fig:f211b}
\end{figure}

\nin We invite the reader to trace the intersections formulae from
Corollary~\ref{crl:6.6} for the example $\tr=(2,1,1)$, shown on
Figure~\ref{fig:f211b}.

\begin{rem}
Corollary~\ref{crl:6.6} implies that every stratum $X_{S,A}$ can be
represented as an intersection of two strata of the type $X_S$, with
only exception provided by the strata $X_{S,S}$, when $|S|=1$.
\end{rem}

\begin{crl}\label{crl:inter}
Assume $S_1,\dots,S_t\subseteq[n]$, such that $S_1\not\subset S_i$,
for all $i=2,\dots,t$. The following two cases describe the
intersection $X_{S_1}\cap\dots\cap X_{S_t}$:
\begin{enumerate}
\item[(1)] if $S_1\supset S_i$, for all $i=2,\dots,t$, then
  $X_{S_1}\cap\dots\cap X_{S_t}=X_{S_1,S_2\cup\dots\cup S_t}$;
\item[(2)] if there exists $2\leq i\leq t$, such that $S_1\not\supset S_i$,
  then $X_{S_1}\cap\dots\cap X_{S_t}=Z_{S_1\cup S_2\cup\dots\cup S_t}
  =X_{S_1\cup S_2\cup\dots\cup S_t,S_1\cup S_2\cup\dots\cup S_t}$.
\end{enumerate}
\end{crl}
\pr Assume first that $S_1\supset S_i$, for all $i=2,\dots,t$. By iterating
\eqref{inc:2} we get
\begin{multline*} X_{S_1}\cap\dots\cap X_{S_t}=X_{S_1,\es}\cap X_{S_2,\es}\cap\dots 
\cap X_{S_t,\es}= X_{S_1,S_2}\cap X_{S_3,\es}\cap\dots\cap
X_{S_t,\es}\\ =X_{S_1,S_2\cup S_3}\cap X_{S_4,\es}\cap\dots\cap
X_{S_t,\es}=\dots=X_{S_1,S_2\cup\dots\cup S_t}.
\end{multline*}
This proves (1). 

To show (2), we can assume without loss of generality, that 
$S_2\not\subset S_1$.  By \eqref{inc:3} we have $X_{S_1}\cap X_{S_2}=Z_{S_1\cup S_2}$.  
By iterating \eqref{eq:xz} we get
\[
Z_{S_1\cup S_2}\cap X_{S_3}\cap\dots\cap X_{S_t}=Z_{S_1\cup S_2\cup
  S_3}\cap X_{S_4}\cap\dots\cap X_{S_t} =X_{S_1\cup S_2\cup\dots\cup
  S_t},
\]
which finishes the proof.
\qed

\subsection{The boundary of the immediate snapshot complexes and its canonical 
decomposition }

\begin{df}\label{df:bv}
Let $\tr$ be an arbitrary round counter, and assume
$V\subset\supp\tr$. We define $B_V(\tr)$ to be the simplicial
subcomplex of $P(\tr)$ consisting of all simplices
$\sigma=((W_0,G_0),\dots,\allowbreak (W_t,G_t))$, satisfying
$V\subseteq G_0$.
\end{df}

The fact that $B_V(\tr)$ is a~well-defined subcomplex of $P(\tr)$ is
immediate from the definition of the ghosting operation. We shall
let $\beta_V(\tr)$ denote the inclusion map
\[\beta_V(\tr):B_V(\tr)\hookrightarrow P(\tr).\]

\begin{prop}
For an arbitrary round counter $\tr$, and any $V\subset\supp\tr$, the map
$\delta_V(\tr)$ given by
\[\delta_V(\tr):((W_0,G_0),\dots,(W_t,G_t))\mapsto((W_0,G_0\sm V),\dots,(W_t,G_t))\]
is a~simplicial isomorphism between simplicial complexes $B_V(\tr)$
and $P(\tr\sm V)$.
\end{prop}
\pr The map $\delta_v(\tr)$ is simplicial, and it has a~simplicial
inverse which adds $V$ to $G_0$.  \qed

\vskip5pt

Given an arbitrary round counter $\tr$, $A\subseteq
S\subseteq\act\tr$, and $V\subset\supp\tr$, such that $S\cap V=\es$,
we set \[X_{S,A,V}(\tr):=X_{S,A}(\tr)\cap B_V(\tr).\] We can use the
notational convention $B_\es(\tr)=P(\tr)$, which is consistent with
Definition~\ref{df:bv}. In this case we get
$X_{S,A,\es}(\tr)=X_{S,A}(\tr)$, fitting well with the previous
notations.

The diagram \eqref{eq:bar} in the next proposition means that we can
naturally think about $X_{S,A,V}(\tr)$ both as $X_{S,A}(\tr\sm V)$ as
well as $B_V(\tr_{S,A})$, or abusing notations we write $B_V\cap
X_{S,A}=X_{S,A}(B_V)=B_V(X_{S,A})$.

\begin{prop}\label{prop:6.13}
Assume $\tr$ is an arbitrary round counter, $V\subset\supp\tr$,
$A\subseteq S\subseteq\act\tr$, and $V\cap S=\es$. Then there exist
simplicial isomorphisms $\varphi$ and $\psi$ making the following
diagram commute:
\begin{equation}\label{eq:bar}
\begin{tikzcd}
P(\tr)\arrow[hookleftarrow]{r}{\alpha}\arrow[hookleftarrow]{d}{\beta} 
&X_{S,A}(\tr)\arrow[squiggly]{r}{\gamma}\arrow[hookleftarrow]{d}{j} 
&P(\tr_{S,A})\arrow[hookleftarrow]{d}{\beta} \\
B_V(\tr)\arrow[hookleftarrow]{r}{i}
\arrow[squiggly]{d}{\delta} 
&X_{S,A,V}(\tr)\arrow[squiggly]{r}{\varphi}\arrow[squiggly]{d}{\psi} 
&B_V(\tr_{S,A})\arrow[squiggly]{d}{\delta} \\
P(\tr\sm V)\arrow[hookleftarrow]{r}{\alpha} 
&X_{S,A}(\tr\sm V)\arrow[squiggly]{r}{\gamma} 
&P(\bar r_{S\cup V,A\cup V}),
\end{tikzcd}
\end{equation}
where $i$ and $j$ denote inclusion maps.
\end{prop}
\pr Note that $X_{S,A,V}(\tr)$ consists of all simplices
$\sigma=((W_0,G_0),\dots,(W_t,G_t))$, such that $V\subseteq G_0$,
$A\subseteq G_1$, and either $W_1\cup G_1=S$, or $S\subseteq G_1$. The
fact that $V$ and $S$ are disjoint ensures that these conditions do
not contradict each other.  We let $\varphi$ be the restriction of
$\gamma_{S,A}(\tr):X_{S,A}(\tr)\ra P(\tr_{S,A})$ to $X_{S,A,V}(\tr)$.
Furthermore, we let $\psi$ be the restriction of
$\delta_V(\tr):B_V(\tr)\ra P(\tr\sm V)$ to $X_{S,A,V}(\tr)$.  \qed

\vskip5pt The commuting diagram \eqref{cd:b1} in the next proposition
shows how the stratum $X_{S,A}(\tr)$ can be naturally interpreted as
a~part of the boundary of the stratum $X_{S,B}(\tr)$, whenever
$B\subseteq A\subseteq S\subseteq\act\tr$.

\begin{prop}\label{prop:5}
Assume $B\subseteq A\subseteq S\subseteq\act\tr$, then the following
diagram commutes
\begin{equation}\label{cd:b1}
\begin{tikzcd}[column sep=large]
X_{S,B}(\tr)\arrow[hookleftarrow]{r}{i}\arrow[squiggly]{d}[swap]{\gamma_{S,B}(\tr)}
& X_{S,A}(\tr)\arrow[squiggly]{rd}{\gamma_{S,A}(\tr)} \\
P(\tr_{S,B})\arrow[hookleftarrow]{r}{\beta_{A\sm B}(\tr_{S,B})}
& B_{A\sm B}(\tr_{S,B})\arrow[squiggly]{r}{\delta_{A\sm B}(\tr_{S,B})}
& P(\tr_{S,A})
\end{tikzcd}
\end{equation}
where $i$ denotes the inclusion map.
\end{prop}
\pr Take $\sigma=((W_0,G_0),\dots,(W_t,G_t))\in X_{S,A}(\tr)$. 
On one hand we have
\[(\gamma_{S,B}(\tr)\circ i)(\sigma)=
\begin{cases}
\begin{array}{|c|c|c|c|}
\hline
W_0\sm G_1       &  W_2 & \dots & W_t \\ \hline
G_0\cup G_1\sm B &  G_2 & \dots & G_t \\ 
\hline
\end{array}, 
\textrm{ if } W_1\cup G_1=S,\,\,A\subseteq G_1;\\[0.6cm]
\begin{array}{|c|c|c|c|c|}
\hline
W_0\sm S  & W_1      & W_2 & \dots & W_t \\ \hline
G_0\cup S\sm B & G_1\sm S & G_2 & \dots & G_t \\ 
\hline
\end{array},
\textrm{ if } S \subseteq G_1.
\end{cases}
\]
On the other hand, we have 
\[(\gamma_{S,A}(\tr))(\sigma)=
\begin{cases}
\begin{array}{|c|c|c|c|}
\hline
W_0\sm G_1       &  W_2 & \dots & W_t \\ \hline
G_0\cup G_1\sm A &  G_2 & \dots & G_t \\ 
\hline
\end{array}, 
\textrm{ if } W_1\cup G_1=S,\,\,A\subseteq G_1;\\[0.6cm]
\begin{array}{|c|c|c|c|c|}
\hline
W_0\sm S  & W_1      & W_2 & \dots & W_t \\ \hline
G_0\cup S\sm A & G_1\sm S & G_2 & \dots & G_t \\ 
\hline
\end{array},
\textrm{ if } S \subseteq G_1.
\end{cases}
\]
Since applying $\delta_{A\sm B}(\tr_{S,B})^{-1}$ will add $A\sm B$ to
  $G_0\cup G_1\sm A$, resp.\ $G_0\cup S\sm A$, above and $A\subseteq
  S$, $A\subseteq G_1$, we conclude that
\[(\gamma_{S,B}(\tr)\circ i)(\sigma)=(\beta_{A\sm B}(\tr_{S,B})\circ
\delta_{A\sm B}(\tr_{S,B})^{-1}\circ\gamma_{S,A}(\tr))(\sigma).\]
Which is the same as to say that the diagram~\eqref{cd:b1} commutes.  
\qed

\subsection{The combinatorial structure of the complexes $P(\chi_{A,B})$} $\,$

\nin Let us analyze the simplicial structure of $P(\chi_{A,B})$. Set
$k:=|A|-1$ and $m:=|B|$. By~\eqref{eq:ptr} the simplicial complex
$P(\chi_{A,B})$ is isomorphic to the $m$-fold suspension of
$P(\chi_A)$. On the other hand, we saw before that $P(\chi_A)$ is
isomorphic to the standard chromatic subdivision of $\Delta^k$. The
simplices of the $m$-fold suspension of $\chi(\da^k)$ (which is of
course homeomorphic to $\da^{m+k}$) are indexed by tuples
$(S,(B_1,\dots,B_t)(C_1,\dots,C_t))$, where $S$ is any subset of $B$,
and the sets $B_1,\dots,B_t,C_1,\dots,C_t$ satisfy the same conditions
as in the combinatorial description of the simplicial structure of
$\chi(\da^k)$.  In line with ~\eqref{eq:bc}, the simplicial
isomorphism between $P(\chi_{A,B})$ and the $m$-fold suspension of
$\chi(\da^k)$ can be explicitly given by
\[(S,(B_1,\dots,B_t)(C_1,\dots,C_t))\mapsto
\begin{array}{|c|c|c|c|}
\hline
W_0            & C_1        & \dots & C_t \\ \hline
(A\cup B)\sm W_0 & B_1\sm C_1 & \dots & B_t\sm C_t \\ 
\hline
\end{array},\]
where $W_0=S\cup B_1\cup\dots\cup B_t$. 
In particular, up to the simplicial
isomorphism, the complex $P(\chi_{A,B})$ depends only on $m$ and~$k$.

The simplices of $P(\chi_{A,B})$ are indexed by all witness structures 
$\sigma=((W_0,G_0),\ab\dots,\ab(W_t,G_t))$ satisfying the following conditions:
\begin{enumerate} 
	\item [(1)] $W_0\cup G_0=A\cup B$;
	\item [(2)] $W_0\cap A=W_1\cup\dots\cup W_t\cup G_1\cup\dots\cup G_t$;
	\item [(3)] the sets $W_1,\dots,W_t,G_1,\dots,G_t$ are disjoint.
\end{enumerate}

It was shown in~\cite{subd} that there is a~homeomorphism
\[\tau_A:P(\chi_A)\,{\underset{\cong}\longrightarrow}\,\Delta^A,\]
such that for any $C\subseteq A$ the following diagram commutes
\begin{equation}\label{cd:tau3}
\begin{tikzcd}[column sep=1.5cm]
P(\chi_A)\arrow[hookleftarrow]{r}{\beta_{A\sm C}(\chi_A)}
\arrow{d}{\cong}[swap]{\tau_A} 
&B_{A\sm C}(\chi_A)\arrow[squiggly]{r}{\delta_{A\sm C}(\chi_A)} 
&P(\chi_C)\arrow{ld}{\cong}[swap]{\tau_C}\\
\da^A\arrow[hookleftarrow]{r}{i}&\da^C
\end{tikzcd}
\end{equation}
where $i:\da^C\hookrightarrow\da^A$ is the standard inclusion map.
In general, given a pair if sets $(A,B)$, we take the $|B|$-fold
suspension of the map $\tau_A$ to produce a~homeomorphism  
\[\tau_{A,B}:P(\chi_{A,B})\underset{\cong}\longrightarrow\Delta^{A\cup B}.\]

\begin{df}\label{df:tau}
When $A\cup B=C\cup D$, we set
\[\tau(\chi_{A,B},\chi_{C,D}):=\tau_{C,D}^{-1}\circ\tau_{A,B},\] 
clearly, we get
a homeomorphism
$\tau(\chi_{A,B},\chi_{C,D}):P(\chi_{A,B})\underset{\cong}\longrightarrow P(\chi_{C,D})$.
\end{df}
We know that this map is a simplicial isomorphism when restricted to
$B_S(\chi_{A,B})$, for all $S\subseteq (A\cap C)\cup (B\cap D)$, i.e.,
we have the following commutative diagram
\begin{equation}
\begin{tikzcd}[column sep=2cm]
B_S(\chi_{A,B})\arrow[squiggly]{r}{\tau(\chi_{A,B},\chi_{C,D})}
\arrow[hookrightarrow]{d}{\beta_S(\chi_{A,B})}  
&B_S(\chi_{C,D})\arrow[hookrightarrow]{d}{\beta_S(\chi_{C,D})}  \\ 
P(\chi_{A,B}) \arrow{r}{\tau(\chi_{A,B},\chi_{C,D})}[swap]{\cong}
&P(\chi_{C,D}) 
\end{tikzcd}
\end{equation}
When $C\subseteq A$, we have $B\subseteq D$, so the condition for $S$
becomes $S\subseteq B\cup C$.  Furthermore, if in addition $T=E\cup
F$, we have 
\[\tau(\chi_{A_1,B_1},\chi_{A_2,B_2})\circ\tau(\chi_{A_2,B_2},\chi_{A_3,B_3})=
\tau(\chi_{A_1,B_1},\chi_{A_3,B_3}).\]

When $A\subseteq C\cup D$, he identity~\eqref{eq:chi2} implies that we
have a~simplicial isomorphism
\[\beta_V(\chi_{C.D}):B_V(\chi_{C,D})\underset\cong\longrightarrow P(\chi_{C\sm A,D\sm A}).\] 
Furthermore, when $S\subseteq C$, the identity~\eqref{eq:chi3} implies
that we have a~simplicial isomorphism
\[X_S(\chi_{C.D}):\gamma_S(\chi_{C,D})\underset\cong\longrightarrow P(\chi_{C\sm S,D\cup S}).\] 

\begin{prop}\label{pr:2.18}
Assume $A\cup B=C\cup D$ and $V\subseteq A\cup B$, then
the following diagram commutes
\begin{equation} \label{cd:tau}
\begin{tikzcd}[column sep=1.5cm]
P(\chi_{A,B})\arrow[hookleftarrow]{r}{\beta_V(\chi_{A,B})}
\arrow{d}{\cong}[swap]{\tau(\chi_{A,B},\chi_{C,D})}
& B_V(\chi_{A,B})\arrow[squiggly]{r}{\delta_V(\chi_{A,B})}
& P(\chi_{A,B}\sm V)\arrow{d}{\tau(\chi_{A,B}\sm V,\chi_{C,D}\sm V)}[swap]{\cong} \\   
P(\chi_{C,D})\arrow[hookleftarrow]{r}{\beta_V(\chi_{C,D})}
& B_V(\chi_{C,D})\arrow[squiggly]{r}{\delta_V(\chi_{C,D})}
& P(\chi_{C,D}\sm V)
\end{tikzcd}
\end{equation}
\end{prop}
\pr Consider the diagram on Figure~\ref{cdf:1}.
\begin{figure}[hbt]
\begin{tikzcd}[column sep=1.5cm]
P(\chi_{A,B})\arrow[hookleftarrow]{r}{\beta_V(\chi_{A,B})}
\arrow{d}{\cong}[swap]{\tau_{A,B}}
& B_V(\chi_{A,B})\arrow[squiggly]{r}{\delta_V(\chi_{A,B})}
& P(\chi_{A,B}\sm V)\arrow{ld}{\tau_{A\sm V,B\sm V}}[swap]{\cong} \\  
\da^{A\cup B}\arrow[hookleftarrow]{r}\arrow[leftarrow]{d}{\cong}[swap]{\tau_{C,D}} 
&\da^{A\cup B\sm V}\arrow[leftarrow]{rd}{\tau_{C\sm V,D\sm V}}[swap]{\cong} \\ 
P(\chi_{C,D})\arrow[hookleftarrow]{r}{\beta_V(\chi_{C,D})}
& B_V(\chi_{C,D})\arrow[squiggly]{r}{\delta_V(\chi_{C,D})}
& P(\chi_{C,D}\sm V)
\end{tikzcd}
\caption{Commuting diagram used in the proof of Proposition~\ref{pr:2.18}}
\label{cdf:1}
\end{figure}
Both the upper and the lower part of this diagram are versions
of~\eqref{cd:tau3}, hence, they commute. Together, they form the
diagram~\eqref{cd:tau}. \qed


\section{Topology of the immediate snapshot complexes}

\subsection{Immediate snapshot complexes are collapsible pseudomanifolds} $\,$

Consider a~quite general situation, where $X$ is an arbitrary
topological space, and $\{X_i\}_{i\in I}$ is a~finite family of
subspace of $X$ covering $X$, that is $I$ is finite and $X=\cup_{i\in
  I}X_i$.

\begin{df}{\rm(\cite[Definition 15.14]{book}).}
The {\bf nerve complex} $\cn$ of a covering $\{X_i\}_{i\in I}$ is
a~simplicial complex whose vertices are indexed by $I$, and a~subset
of vertices $J\subseteq$ spans a~simplex if and only if the
intersection $\cap_{i\in J}X_i$ is not empty.
\end{df}

The nerve complex can be useful because of the following fact.
\begin{lm} \label{lm:nerve}
{\rm(Nerve Lemma, \cite[Theorem 15.21, Remark 15.22]{book}).}  Assume
$K$ is a simplicial complex, covered by a~family of subcomplexes
$\ck=\{K_i\}_{i\in I}$, such that $\cap_{i\in J}K_i$ is empty or
contractible for all $J\subseteq I$, then $K$ is homotopy equivalent
to the nerve complex $\cn(\ck)$.
\end{lm}

\begin{crl}
For an arbitrary round counter $\tr$, the simplicial complex $P(\tr)$
is contractible.
\end{crl}
\pr We use induction on $|\tr|$. If $|\tr|=0$, then $P(\tr)$ is just a
simplex, hence contractible. We assume that $|\tr|\geq 1$, and view
the canonical decomposition $P(\tr)=\cup_{S\subseteq\act\tr} X_S(\tr)$
as a~covering of $P(\tr)$.  By Proposition~\ref{prop:strata},
Corollary~\ref{crl:inter}, and the induction assumption, all the
intersections of the subcomplexes $X_S(\tr)$ with each other are
either empty or contractible. This means, that we can apply the Nerve
Lemma~\ref{lm:nerve}, with $K=P(\tr)$, $I=2^{\act\tr}\sm\{\es\}$, and
$K_i$'s are $X_S(\tr)$'s.

Now, by Corollary~\ref{crl:inter} we see that $X_{\act\tr}\cap
X_S=X_{\act\tr,S}\neq\es$ for all $S\subset\act\tr$. It follows that
the nerve complex of this decomposition as a~cone with apex at
$\act\tr\in I$. Since the nerve complex is contractible, it follows
from the Nerve Lemma~\ref{lm:nerve} that $P(\tr)$ is contractible as
well.  \qed

While contractibility is a property of topological spaces, there is
a~stronger combinatorial property called {\it collapsibility},
see~\cite{coh}, which some simplicial complexes may have.

\begin{df}
Let $K$ be a simplicial complex. A pair of simplices $(\sigma,\tau)$
of $K$ is called an~{\bf elementary collapse} if the following
conditions are satisfied:
\begin{itemize}
\item $\tau$ is a maximal simplex,
\item $\tau$ is the only simplex which properly contains $\sigma$.
\end{itemize}
A finite simplicial complex $K$ is called {\bf collapsible}, if there
exists a sequence $(\sigma_1,\tau_1),\dots,\allowbreak
(\sigma_t,\tau_t)$ of pairs of simplices of $K$, such that
\begin{itemize}
\item this sequence yields a perfect matching on the set of all simplices of~$K$,
\item for every $1\leq k\leq t$, the pair $(\sigma_k,\tau_k)$ is an
  elementary collapse in
  $K\sm\{\sigma_1,\dots,\sigma_{k-1},\allowbreak\tau_1,\dots,\tau_{k-1}\}$.
\end{itemize}
\end{df}

When $(\sigma,\tau)$ is an elementary collapse, we also say that $\sigma$
is a~{\it free} simplex.

We have shown in Proposition~\ref{prop:pseudo} that for any round counter $\tr$ 
the simplicial complex $P(\tr)$ is a pseudomanifold with boundary $\partial P(\tr)$.
Set 
\[\inte P(\tr):=\bigcup_{\sigma\in P(\tr),\,\,\sigma\notin\partial P(\tr)}\inte\sigma,\]
and, for all $A\subseteq S\subseteq\act\tr$, set
\[\partial X_{S,A}(\tr):=\gamma_{S,A}(\tr)^{-1}(\partial P(\tr_{S,A})),\quad
\inte X_{S,A}(\tr):=\gamma_{S,A}(\tr)^{-1}(\inte P(\tr_{S,A})).\]

\begin{prop}\label{prop:6.9}
Assume $\tr$ is an arbitrary round counter, $A\subset S
\subseteq\act\tr$, and $V\subseteq\supp\tr\sm S$. The simplicial
complex $\partial X_{S,A,V}(\tr)$ is the subcomplex of
$X_{S,A,V}(\tr)$ consisting of all simplices
\[\sigma=((W_0,V),\allowbreak (S\sm A,A),\dots,(W_t,G_t)).\]
\end{prop}
\pr Pick $\sigma\in X_{S,A,V}$, and set $\rho$ to be the composition
of the simplicial isomorphisms $X_{S,A,V}(\tr)\ra B_V(\tr_{S,A})\ra
P(\tr_{S\cup V, A\cup V})$ from the commutative
diagram~\eqref{eq:bar}. 

Assume first that $W_1\cup G_1=S$, then
\[\rho(\sigma)=((W_0\sm G_1,(G_0\cup G_1)\sm(A\cup V)),(W_2,G_2),
\dots,(W_t,G_t)).\]
Clearly $\rho(\sigma)\notin\partial P(\tr_{S\cup V,A\cup V})$ if and
only if $(G_0\cup G_1)\sm(A\cup V)=\es$, i.e., $G_0\cup G_1\subseteq
A\cup V$. Since we know that $A\subseteq G_1$, $V\subseteq G_0$, this
means that $G_0=V$ and $G_1=A$, which implies $W_1=S\sm A$. 

Assume now that $S\subseteq G_1$, then we have 
\[\rho(\sigma)=((W_0\sm S,(G_0\cup S)\sm(A\cup V)),(W_1,G_1\sm S),
(W_2,G_2),\dots,(W_t,G_t)).\] Here we have $\rho(\sigma)\notin\partial
P(\tr_{S\cup V,A\cup V})$ if and only if $(G_0\cup S)\sm(A\cup V)
=\es$, which is impossible, since $V\cap S=\es$, and $A\subset V$.
\qed

\begin{crl}\label{crl:strata}
The simplicial complex $P(\tr)$ can be decomposed as a disjoint union
of the simplex $\da^{\pass\tr}=((\pass\tr,\act\tr))$, and the sets
$\inte X_{S,A,V}$, where $(S,A,V)$ range over all triples satisfying
$A\subset S\subseteq\act\tr$ and $V\subseteq\supp\tr\sm S$.

Specifically, for a simplex $\sigma\in P(\tr)$,
$\sigma=((W_0,G_0),\dots,(W_t,G_t))$, we have: if $t=0$, then
$\sigma\subseteq\Delta^{\pass\tr}$, else $\inte\sigma\subseteq\inte
X_{W_1\sm G_1,G_1,G_0}$.
\end{crl}
\pr Immediate from Proposition~\ref{prop:6.9}.
\qed

\begin{lm}\label{lm:coll}
Assume $\tr$ is a round counter, and $p\in\supp\tr$, then there exists
a sequence of elementary collapses reducing the simplicial complex
$P(\tr)$ to the subcomplex $(\partial P(\tr))\sm\inte B_p(\tr)$.
\end{lm}
\pr The proof is again by induction on $|r|$. The case $|r|=0$ is
trivial. The simplices we need to collapse are precisely those, whose
interior lies in $\inte P(\tr)\cup\inte B_p(\tr)$. Let $\Sigma$ denote
the set of all strata $X_{S,A}$, where $A\subset S\subseteq\act\tr$,
together with all strata $X_{S,A,p}$, where $A\subset
S\subseteq\act\tr$, $p\notin S$. By Corollary~\ref{crl:strata}, the
union of the interiors of the strata in $\Sigma$ is precisely $\inte
P(\tr)\cup\inte B_p(\tr)$.

We describe our collapsing as a~sequence of steps. At each step we
pick a~certain pair of strata $(Y,X)$, where $Y\subset X$, which we
must ``collapse''. Then, we use one of the previous results to show
that as a~simplicial pair $(Y,X)$ is isomorphic to
$(B_t(\tr'),P(\tr'))$, for some round counter $\tr'$, such that
$|\tr'|<|\tr|$. By induction assumption this means that there is
a~sequence of simplicial collapses which removes $\inte X\cup\inte Y$.
Finally, we order these pairs of strata with disjoint interiors
$(Y_1,X_1),\dots,(Y_d,X_d)$ such that for every $1\leq i\leq d$, every
simplex $\sigma\in P(\tr)$, such that $\inte\sigma\subseteq\inte
X_i\cup\inte Y_i$, and every $\tau\supset\sigma$, such that
$\dim\tau=\dim\sigma+1$, we have
\begin{equation}\label{eq:intt}
\inte\tau\subseteq\inte X_1\cup\dots\cup\inte X_i\cup
\inte Y_1\cup\dots\cup\inte Y_i.
\end{equation}
This means, that at step $i$ we can
collapse away the pair of strata $(Y_i,X_i)$ (i.e., collapse away
those simplices whose interior is contained in $\inte X_i\cup\inte
Y_i$) using the procedure given by the induction assumption, and that
these elementary collapses will be legal in $P(\tr)\sm
(X_1\cup\dots\cup\inte X_{i-1}\cup\inte Y_1\cup\dots\cup\inte
Y_{i-1})$ as well.

Our procedure is now divided into $3$ stages.  At stage 1, we match
the strata $X_{S,A,p}$ with $X_{S,A}$, for all $A\subset S
\subseteq\act\tr$, such that $p\notin S$. It follows from the
commutativity of the diagram~\eqref{eq:bar} that each pair of
simplicial subcomplexes $(X_{S,A,p},X_{S,A})$ is isomorphic to the
pair $(B_p(\tr_{S,A}),P(\tr_{S,A}))$. We have
$|\tr_{S,A}|\leq|\tr|-|S|<|\tr|$, hence by induction assumption, this
pair can be collapsed. As a collapsing order we choose any order which
does not decrease the cardinality of the set $A$. Take $\sigma$ such
that $\inte\sigma\subseteq\inte X_{S,A,p}\cup\inte X_{S,A}$. By
Proposition~\ref{prop:6.9} this means that $\sigma=((W_0,T),(S\sm A,A),\dots)$,
where either $T=\es$, or $T=\{p\}$. Take $\tau\supset\sigma$,
such that $\dim\tau=\dim\sigma+1$. Then by Proposition~\ref{prop:b}(b) there exists
$q\in A(\tau)$, such that $\sigma=\Gamma_q(\tau)$. 
A~case-by-case analysis of the ghosting construction shows that 
$\inte\tau\subseteq\inte X$, where $X$ is one of the following strata:
$X_{S,A}$, $X_{S,A,p}$, $X_q$, $X_{q,\es,p}$, $X_{S,A\sm\{q\}}$, 
$X_{S,A\sm\{q\},p}$. Since the order in which we do collapses
does not decrease the cardinality of $A$, the interiors of the last $4$ of 
these strata have already been removed, hence the condition~\eqref{eq:intt}
is satisfied.

At stage 2, we match $X_S$ with $X_{S,S\sm\{p\}}$, for all
$S\subseteq\act\tr$, such that $p\in S$, $|S|\geq 2$. By commutativity of the 
diagram~\eqref{cd:b1}, the pair $(X_{S,S\sm\{p\}},X_S)$ is isomorphic to
$(B_{S\sm\{p\}}(\tr_S),P(\tr_S))$. This big collapse can easily be
expressed as a sequence of elementary collapses, though in
a~non-canonical way.  For this, we pick any $q\in S\sm\{p\}$. It
exists, since we assumed that $|S|\geq 2$.  Then we match pairs
$(X_{S,A\cup\{q\}},X_{S,A})$, for all $A\subseteq S\sm\{q\}$.  
Again, by commutativity of the diagram~\eqref{cd:b1}, this pair 
is isomorphic to $(B_q(\tr_{S,A}),P(\tr_{S,A}))$. The
order in which we arrange $S$ does not matter for the collapsing
order.  Once $S$ is fixed, the collapsing order inside does not
decrease the cardinality of $A$. As above, take $\sigma$ such
that $\inte\sigma\subseteq\inte X_{S,A\cup\{q\}}\cup\inte X_{S,A}$,
take $\tau\supset\sigma$, such that $\dim\tau=\dim\sigma+1$, and take
$r\in A(\tau)$, such that $\sigma=\Gamma_q(\tau)$. By
Proposition~\ref{prop:6.9} we have $\sigma=((W_0,\es),(S\sm A,A),\dots)$,
or $\sigma=((W_0,\es),(S\sm(A\cup\{q\}),A\cup\{q\}),\dots)$. Note,
that both $q$ and $r$ are different from $p$, but we may have $q=r$.
Again, a~case-by-case analysis of the ghosting construction shows that 
$\inte\tau\subseteq\inte X$, where $X$ is one of the following strata:
$X_{S,A}$, $X_{S,A\cup\{q\}}$, $X_{S,A\sm\{r\}}$, $X_{S,A\cup\{q\}\sm\{r\}}$, 
$X_q$, $X_r$. Again, since collapsing order does not decrease the cardinality
of~$A$, the condition~\eqref{eq:intt} is satisfied.

At stage 3, we collapse the pair $(X_{p,p},X_p)$. Let us be specific. 
First, by Corollary~\ref{crl:xaa} we know that 
$X_{p,p}=\bigcup_{\{p\}\subset S\subseteq\act\tr}X_{S,p}$, and 
it follows from Proposition~\ref{prop:6.9} that 
$\inte X_{p,p}=\bigcup_{\{p\}\subset S\subseteq\act\tr}\inte X_{S,p}$. 
By commutativity of the diagram~\eqref{cd:b1}, the pair $(X_{p,p},X_p)$
is isomorphic to $(B_p(\tr_p),P(\tr_p))$, hence it can be collapsed
using the induction assumption. Clearly, the entire procedure exhausts
the set $\Sigma$, and we arrive at the simplicial complex 
$(\partial P(\tr))\sm\inte B_p(\tr)$.
\qed

\begin{crl}
For an arbitrary round counter $\tr$, the simplicial complex $P(\tr)$
is collapsible.
\end{crl}
\pr Iterative use of Lemma~\ref{lm:coll}. 
\qed

\subsection{Homeomorphic gluing}

\begin{df}
We say that a simplicial complex $K$ is {\bf simplicially homeomorphic}
to a~simplex $\Delta^A$, where $A$ is some finite set, if there exists 
a~homeomorphism $\varphi:\da^A\ra K$, such that for every simplex 
$\sigma\in\da^A$, the image $\varphi(\sigma)$ is a~subcomplex of~$K$.
\end{df}

When we say that a CW complex is {\it finite} we shall mean that it has
finitely many cells.

\begin{df}\label{df:gd}
 Let $X$ and $Y$ be finite CW complexes. A~{\bf homeomorphic gluing
   data} between $X$ and $Y$ consists of the following:
\begin{itemize}
\item a~family $(A_i)_{i=1}^t$ of CW subcomplexes of $X$, such that
  $X=\cup_{i=1}^t A_i$,
\item a~family $(B_i)_{i=1}^t$ of CW subcomplexes of $Y$, such that
  $Y=\cup_{i=1}^t B_i$,
\item a~family of homeomorphisms $(\varphi_i)_{i=1}^t$,
  $\varphi_i:A_i\rightarrow B_i$,
\end{itemize}
satisfying the compatibility condition: if $x\in A_i\cap A_j$, then
$\varphi_i(x)=\varphi_j(x)$.
\end{df}

Given finite CW complexes $X$ and $Y$, together with homeomorphic
gluing data $(A_i,B_i,\varphi_i)_{i=1}^t$ from $X$ to $Y$, we define
$\varphi:X\rightarrow Y$, by setting $\varphi(x):=\varphi_i(x)$,
whenever $x\in A_i$. The compatibility condition from
Definition~\ref{df:gd} implies that $\varphi(x)$ is independent of the
choice of $i$, hence the map $\varphi:X\rightarrow Y$ is well-defined.

\begin{lm} \label{lm:hg} {\rm (Homeomorphism Gluing Lemma).}

\noindent Assume we are given finite CW complexes $X$ and $Y$, and
homeomorphic gluing data $(A_i,B_i,\varphi_i)_{i=1}^t$, satisfying an
additional condition:
\begin{equation} \label{eq:c2}
\textrm{ if } \varphi(x)\in B_i, \textrm{ then } x\in A_i,
\end{equation}
then the map $\varphi:X\rightarrow Y$ is a~homeomorphism.
\end{lm}
\pr First it is easy to see that $\varphi$ is surjective. 
Take an arbitrary $y\in Y$, then there exists $i$ such that $y\in B_i$.
Take $x=\varphi_i^{-1}(y)$, clearly $\varphi(x)=y$.

Let us now check the injectivity of $\varphi$. Take $x_1,x_2\in X$
such that $\varphi(x_1)=\varphi(x_2)$. There exists $i$ such that
$x_1\in A_i$.  Then $\varphi(x_1)=\varphi_i(x_1)\in B_i$, hence
$\varphi(x_2)\in B_i$.  Condition~\eqref{eq:c2} implies that $x_2\in
A_i$. The fact that $x_1=x_2$ now follows from the injectivity of
$\varphi_i$.

We have verified that $\varphi$ is bijective, so
$\varphi^{-1}:Y\rightarrow X$ is a well-defined map. We shall now
prove that $\varphi^{-1}$ is continuous by showing that $\varphi$
takes closed sets to closed sets. To start with, let us recall the
following basic property of the topology of CW complexes: a subset $A$
of a CW complex $X$ is closed if and only if its intersection with the
closure of each cell in $X$ is closed. Sometimes, one uses the
terminology {\it weak topology} of the CW complex. This property was
an integral part of the original J.H.C.\ Whitehead definition of CW
complexes, see, e.g.,~\cite[Proposition A.2.]{Hat} for further
details.

Let us return to our situation. We claim that $A\subseteq X$ is
closed, if and only if $A\cap A_i$ is closed in $A_i$, for each
$i=1,\dots,t$. Note first that since $A_i$ is itself closed, a~subset
$S\subseteq A_i$ is closed in $X$ if and only if it is closed in
$A_i$, so we will skip mentioning where the sets are closed.  Clearly,
if $A$ is closed, then $A\cap A_i$ is closed for all~$i=1,\dots,t$.
On the other hand, assume $A\cap A_i$ is closed for all~$i$. Let
$\sigma$ be a closed cell of $X$, we need to show that $A\cap\sigma$
is closed.  Since $X=\cup_{i=1}^t A_i$, and $A_i$'s are CW
subcomplexes of $X$, there exists $i$, such that $\sigma\subseteq
A_i$. Then $A\cap\sigma=A\cap (A_i\cap\sigma)=(A\cap A_i)\cap\sigma$,
but $(A\cap A_i)\cap\sigma$ is closed since $A\cap A_i$ is closed.
Hence $A\cap\sigma$ is closed and our argument is finished.
Similarly, we can show that $B\subseteq X$ is closed, if and only if
$B\cap B_i$ is closed, for each $i=1,\dots,t$.
 
Pick now a~closed set $A\subseteq X$, we want to show that
$\varphi(A)$ is closed.  To start with, for all $i$ the set $A\cap
A_i$ is closed, hence $\varphi_i(A\cap A_i)\subseteq B_i$ is also
closed, since $\varphi_i$ is a~homeomorphism.  Let us verify that for
all $i$ we have
\begin{equation}\label{eq:jhc}
\varphi_i(A\cap A_i)=\varphi(A)\cap B_i.
\end{equation} 
Assume $y\in\varphi_i(A\cap A_i)$. On one hand $y\in\varphi_i(A_i)$,
so $y\in B_i$, on the other hand, $y=\varphi_i(x)$, for $x\in A$, so
$y\in\varphi(A)$. Reversely, assume $y\in\varphi(A)$ and $y\in
B_i$. Then $y=\varphi(x)\in B_i$, so condition~\eqref{eq:c2} implies
that $x\in A_i$, hence $y\in\varphi(A\cap A_i)$, which
proves~\eqref{eq:jhc}. It follows that $\varphi(A)\cap B_i$ is closed
for all $i$, hence $\varphi(A)$ itself is closed. This proves that
$\varphi^{-1}$ is continuous.

We have now shown that $\varphi^{-1}:Y\rightarrow X$ is a~continuous
bijection.  Since $X$ and $Y$ are both finite CW complexes, they are
compact Hausdorff when viewed as topological spaces.  It is a~basic
fact of set-theoretic topology that a~continuous bijection between
compact Hausdorff topological spaces is automatically a~homeomorphism,
see e.g., \cite[Theorem 26.6]{Mun}.  \qed

\vspace{5pt}

The following variations of the Homeomorphism Gluing Lemma~\ref{lm:hg}
will be useful for us. 

\begin{crl} \label{crl:hg}
 Assume we are given finite CW complexes $X$ and $Y$, and
homeomorphic gluing data $(A_i,B_i,\varphi_i)_{i=1}^t$, satisfying an
additional condition:
\begin{equation} \label{eq:c3}
\textrm{ for all } I\subseteq[t]: \varphi:A_I\rightarrow B_I
\textrm{ is a~bijection. }
\end{equation}
Then the map $\varphi:X\rightarrow Y$ is a~homeomorphism.
\end{crl}
\pr Clearly, we just need to show that the condition \eqref{eq:c3}
implies the condition~\eqref{eq:c2}. Assume $y=\varphi(x)$, $y\in
B_i$, and $x\not\in A_i$. Let $I$ be the maximal set such that $y\in
B_I$. The condition~\eqref{eq:c3} implies that there exists a~unique
element $\tilde x\in A_I$, such that $\varphi(\tilde x)=y$. In
particular, $\tilde x\in A_i$, hence $x\neq\tilde x$. Even stronger,
if $x\in A_i$, for some $i\in I$, then $x,\tilde x\in A_i$, hence
$x=\tilde x$, since $\varphi_i$ is injective. So $x_i\not\in A_i$, for
all $i\in I$. Hence, there exists $j\not\in I$, such that $x\in A_j$,
which implies $\varphi(x)\in B_j$, yielding a~contradiction to the
maximality of the set~$I$. \qed

\begin{crl}\label{crl:hg2}
Assume we are given CW complexes $X$ and $Y$, a collection
$(A_i)_{i=1}^t$ of CW subcomplexes of $X$, a collection
$(B_i)_{i=1}^t$ of CW subcomplexes of $Y$, and a collection
$(\varphi_I)_{I\subseteq[t]}$ of maps such that 
\begin{itemize}
\item $X=\cup_{i=1}^t A_i$, $Y=\cup_{i=1}^t B_i$; 
\item for every $I\subseteq[t]$, the map $\varphi_I:A_I\rightarrow
  B_I$ is a~homeomorphism;
\item for every $J\supseteq I$ the following diagram commutes
\begin{equation} \label{cd:ij1}
\begin{tikzcd}
A_J\arrow{r}{\varphi_J}[swap]{\cong}\arrow[hookrightarrow]{d} 
&B_J\arrow[hookrightarrow]{d} \\
A_I \arrow{r}{\varphi_I}[swap]{\cong} &B_I
\end{tikzcd}
\end{equation}
\end{itemize}
Then $(A_i,B_i,\varphi_i)_{i=1}^t$ is a homeomorphic gluing data, and
the map $\varphi:X\rightarrow Y$ defined by this data is
a~homeomorphism.
\end{crl}
\pr For arbitrary $1\leq i,j\leq t$, commutativity of \eqref{cd:ij1}
implies that also the following diagram is commutative
\[\begin{tikzcd}
A_i\arrow[hookleftarrow]{r}\arrow{d}{\varphi_{\{i\}}}[swap]{\cong} 
&A_{\{i,j\}}\arrow[hookrightarrow]{r}\arrow{d}{\varphi_{\{i,j\}}}[swap]{\cong} 
&A_j\arrow{d}{\varphi_{\{j\}}}[swap]{\cong} \\
B_i\arrow[hookleftarrow]{r} 
&B_{\{i,j\}}\arrow[hookrightarrow]{r} &B_j
\end{tikzcd}\]
In other words, for any $x\in A_i\cap A_j$, we have
$\varphi_{\{i\}}(x)=\varphi_{\{i,j\}}(x)=\varphi_{\{j\}}(x)$.  It
follows that $(A_i,B_i,\varphi_{\{i\}})_{i=1}^t$ is a~homeomorphic
gluing data. Since for all $I\subseteq[t]$, the map $\varphi_I$ is
a~homeomorphism, it is in particular bijective, so conditions of
Corollary~\ref{crl:hg} are satisfied, and the defined map $\varphi$ is
a~homeomorphism. \qed


\subsection{Main Theorem}

The fact that the protocol complexes in the immediate snapshot
read/write shared memory model are homeomorphic to simplices has
been folklore knowledge in the theoretical distributed computing
community, \cite{Herl}. The next theorem provides a rigorous
mathematical proof of this fact.

\begin{thm}\label{thm:main}

For every round counter $\tr$ there exists a~homeomorphism 
\[\Phi(\tr):P(\tr)\stackrel\cong\longrightarrow P(\chi(\tr)),\]
such that 
\begin{enumerate}
\item[(1)] for all $V\subset\supp\tr$ the following diagram commutes:
\begin{equation}\label{cd:b}
\begin{tikzcd}[column sep=1.1cm]
P(\tr\sm V)\arrow[leftsquigarrow]{r}{\delta_V(\tr)} 
\arrow{d}{\cong}[swap]{\Phi(\tr\sm V)}
&B_V(\tr)\arrow[hookrightarrow]{r}{\beta_V(\tr)} 
& P(\tr)\arrow{d}{\Phi(\tr)}[swap]{\cong} \\ 
P(\chi(\tr\sm V))\arrow[leftsquigarrow]{r}{\delta_V(\chi(\tr))} 
& B_V(\chi(\tr))\arrow[hookrightarrow]{r}{\beta_V(\chi(\tr))} 
&P(\chi(\tr)) 
\end{tikzcd}
\end{equation}
\item[(2)]  for all $S\subseteq\act\tr$ the following diagram commutes:
\begin{equation}\label{cd:b2}
\begin{tikzcd}[column sep=1.1cm]
X_S(\tr)\arrow[squiggly]{r}{\gamma_S(\tr)}\arrow[hookrightarrow]{rd}{\alpha_S(\tr)}
&P(\tr_S)\arrow{r}{\Phi(\tr_S)}[swap]{\cong}
&P(\chi(\tr_S))\arrow{r}{\tau}[swap]{\cong}
&P(\chi(\tr)_S)\arrow[leftsquigarrow]{r}{\gamma_S(\chi(\tr))}
&X_S(\chi(\tr))\\
&P(\tr)\arrow{rr}{\Phi(\tr)}[swap]{\cong}
& & P(\chi(\tr))\arrow[hookleftarrow]{ru}{\alpha_S(\chi(\tr))}
\end{tikzcd}
\end{equation}
where $\tau=\tau(\chi(\tr_S),\chi(\tr)_S)$.
\end{enumerate}
In particular, the complex $P(\tr)$ is simplicially homeomorphic to $\da^{\supp\tr}$.
\end{thm}

\pr Our proof is a~double induction, first on $|\supp\tr|$, then, once
$|\supp\tr|$ is fixed, on the cardinality of the round counter
$\tr$. As a~base of the induction, we note that the case
$|\supp\tr|=1$ is trivial, since the involved spaces are
points. Furthermore, if $|\supp\tr|$ is fixed, and $|\tr|=0$, we take
$\Phi(\tr)$ to be the identity map. In this case the simplicial
complexes $P(\tr)$ and $P(\chi(\tr))$ are simplices. The
diagram~\eqref{cd:b} commutes, since also $\Phi(\tr\sm V)$ is the
identity map. The condition (2) of the theorem is void, since
$\act\tr=\es$. As a~matter of fact, more generally, $\Phi(\tr)$ can be
taken to be the identity map whenever $\tr=\chi(\tr)$, that is
whenever $\tr(i)\in\{0,1\}$, for all $i\in\supp\tr$.

We now proceed to prove the induction step, assuming that $|\tr|\geq
1$.  For every pair of sets $A\subseteq S\subseteq\act\tr$, such that
$S\neq\es$, we define a~map
\[\varphi_{S,A}(\tr):X_{S,A}(\tr)\longrightarrow X_{S,A}(\chi(\tr)),\]
as follows
\begin{equation} \label{cd:phi}
\begin{tikzcd}[column sep=0.9cm]
\varphi_{S,A}(\tr):X_{S,A}(\tr)\arrow[squiggly]{r}{\gamma_{S,A}(\tr)} 
&P(\tr_{S,A}) \arrow{r}{\Phi(\tr_{S,A})}[swap]{\cong} 
&P(\chi(\tr_{S,A}))\arrow{r}{\tau}[swap]{\cong} 
&P(\chi(\tr)_{S,A})\arrow[leftsquigarrow]{r}{\gamma_{S,A}(\chi(\tr))} 
&X_{S,A}(\chi(\tr)),
\end{tikzcd}
\end{equation}
where $\tau=\tau(\chi(\tr_{S,A}),\chi(\tr)_{S,A})$. Since
$|\tr_{S,A}|\leq|\tr_S|=|\tr|-|S|<|\tr|$, the map $\Phi(\tr_{S,A})$ is
already defined by induction, so $\varphi_{S,A}(\tr)$ is well-defined
by the sequence~\eqref{cd:phi}. Obviously, the map $\varphi_{S,A}$ is
a~homeomorphism for all pairs $S,A$.

We want to use Corollary~\ref{crl:hg2} to construct the global
homeomorphism $\Phi(\tr)$ by gluing the local ones
$\varphi_{S,A}(\tr)$. In our setting here, the notations of
Corollary~\ref{crl:hg2} translate to $X=P(\tr)$, $Y=P(\chi(\tr))$,
$A_I$'s are $X_{S,A}(\tr)$'s, $B_I$'s are $X_{S,A}(\chi(\tr))$'s, and
$\varphi_I$'s are $\varphi_{S,A}(\tr)$'s. To satisfy the conditions of
Corollary~\ref{crl:hg2}, we need to check that the following diagram
commutes whenever $X_{S,A}\subseteq X_{T,B}$
\begin{equation}\label{cd:m1}
\begin{tikzcd}
X_{S,A}(\tr)\arrow{r}{\varphi_{S,A}(\tr)}[swap]{\cong}
\arrow[hookrightarrow]{d}{i} 
&X_{S,A}(\chi(\tr))\arrow[hookrightarrow]{d}{j}  \\ 
X_{T,B}(\tr)\arrow{r}{\varphi_{T,B}(\tr)}[swap]{\cong} 
&X_{T,B}(\chi(\tr)),  
\end{tikzcd}
\end{equation}
where $i$ and $j$ denote the inclusion maps.

Note, that by Proposition~\ref{pr:inc1}, we have $X_{S,A}\subseteq
X_{T,B}$ if and only if either $S=T$ and $B\subseteq A$, or
$T\subseteq A$.  Consider first the case $S=T$, $B=\es$. Consider the
diagram on Figure~\ref{cdf:m2}.
\begin{figure}[hbt]
\begin{tikzcd}
\,& P(\tr_{S,A})\arrow{r}{\Phi}[swap]{\cong} 
&P(\chi(\tr_{S,A}))\arrow{r}{\tau}[swap]{\cong} 
&P(\chi(\tr)_{S,A}) \\
X_{S,A}(\tr)\arrow[squiggly]{ru}{\gamma}\arrow[hookrightarrow]{d} 
&B_A(\tr_S)\arrow[squiggly]{u}[swap]{\delta}\arrow[hookrightarrow]{d}{\beta} 
&B_A(\chi(\tr_S))\arrow[squiggly]{u}[swap]{\delta}\arrow[hookrightarrow]{d}{\beta} 
&B_A(\chi(\tr)_S)\arrow[hookrightarrow]{d}{\beta}\arrow[squiggly]{u}[swap]{\delta}
&X_{S,A}(\chi(\tr))\arrow{d}\arrow[squiggly]{lu}{\gamma} \\
X_S(\tr)\arrow[squiggly]{r}{\gamma} 
&P(\tr_S)\arrow{r}{\Phi}[swap]{\cong} 
&P(\chi(\tr_S))\arrow{r}{\tau}[swap]{\cong} &P(\chi(\tr)_S) 
&X_S(\chi(\tr))\arrow[squiggly]{l}{\gamma}
\end{tikzcd}
\caption{Commuting diagram used in the proof of Theorem~\ref{thm:main}}
\label{cdf:m2}
\end{figure}
The leftmost pentagon is the diagram \eqref{cd:b1}, which commutes by
Proposition~\ref{prop:5}. The following hexagon is the diagram
\eqref{cd:b}, where $\tr$ is replaced with $\tr_S$. Since
$|\tr_S|=|\tr|-|S|<|\tr|$, this diagram commutes by induction. The
next hexagon is the diagram \eqref{cd:tau}, for
$\chi_{C_1,D_1}=\chi(\tr_S)$, $\chi_{C_2,D_2}=\chi(\tr)_S$, and we use
the fact that $\chi(\tr_S)\sm A=\chi(\tr_{S,A})$. Finally, the
rightmost pentagon is also the commuting diagram \eqref{cd:b1}, where
$\tr$ is replaced with $\chi(\tr)$. Since removing the $3$ inner terms
of the diagram on Figure \ref{cdf:m2} yields the diagram~\eqref{cd:m1} with
$S=T$, $B=\es$, we conclude that \eqref{cd:m1} commutes in this
special case.

Consider now the case $S=T$, $B\subseteq A$. We have inclusions
$X_{S,A}\hra X_{S,B}\hra X_{S}$, and it is easy to see that the
commutativity of the diagram \eqref{cd:m1} for the inclusion
$X_{S,A}\hra X_{S,B}$ follows from the commutativity of the diagrams
\eqref{cd:m1} for the inclusions $X_{S,A}\hra X_S$ and $X_{S,B}\hra
X_S$. Hence we are done with the proof of this case.

Let us now prove the commutativity of the diagram~\eqref{cd:m1} for
the inclusion $X_{S,A}\hra X_{T,B}$, when $T\subseteq A$.
Assume first that $A=T=B\neq\es$, and consider the diagram on Figure~\ref{cdf:mt1}, where $\wti S=S\sm A$.

\begin{figure}[hbt]
\begin{tikzcd}[column sep=0.6cm]
\, &P(\tr_{S,A})\arrow{r}{\Phi}[swap]{\cong} 
&P(\chi(\tr_{S,A}))\arrow{r}{\tau}[swap]{\cong} 
& P(\chi(\tr)_{S,A}) \\
X_{S,A}(\tr)\arrow[hookrightarrow]{d}\arrow[squiggly]{ru}{\gamma} 
&X_{\wti S}(\tr\sm A)\arrow[hookrightarrow]{d}{\alpha}\arrow[squiggly]{u}[swap]{\gamma}  &  
&X_{\wti S}(\chi(\tr\sm A))\arrow[hookrightarrow]{d}{\alpha}\arrow[squiggly]{u}[swap]{\gamma} 
&X_{S,A}(\chi(\tr))\arrow[hookrightarrow]{d}\arrow[squiggly]{lu}[swap]{\gamma}  \\
X_{A,A}(\tr)\arrow[squiggly]{r}{\gamma} 
&P(\tr\sm A)\arrow{rr}{\Phi}[swap]{\cong}  & 
&P(\chi(\tr\sm A))
&X_{A,A}(\chi(\tr))\arrow[squiggly]{l}[swap]{\gamma},
\end{tikzcd}
\caption{Commuting diagram used in the proof of Theorem~\ref{thm:main}}
\label{cdf:mt1}
\end{figure}
 A few of the maps in the diagram on Figure~\ref{cdf:mt1} 
need to be articulated.  To start with, we have the identity $\tr_{A,A}=\tr\sm
A$, explaining the simplicial isomorphism
$\gamma_{A,A}(\tr):X_{A,A}(\tr)\ra P(\tr\sm A)$. Similarly,
$\chi(\tr)_{A,A}=\chi(\tr\sm A)$ explains
$\gamma_{A,A}(\chi(\tr)):X_{A,A}(\chi(\tr))\ra P(\chi(\tr\sm A))$.
Furthermore, by~\eqref{eq:rsa1} we have $\tr\sm A\dar\wti S=\tr\dar
S\sm A=\tr_{S,A}$ and $\chi(\tr\sm A)\dar\wti S=\chi(\tr)\sm A\dar\wti
S=\chi(\tr)_{S,A}$. These identities explain the presence of the maps
$\gamma_{\wti S}(\tr\sm A):X_{\wti S}(\tr\sm A)\ra P(\tr_{S,A})$, and
$\gamma_{\wti S}(\chi(\tr\sm A)):X_{\wti S}(\chi(\tr\sm A))\ra
P(\chi(\tr)_{S,A})$.

Let us look at the commutativity of the diagram on Figure~\ref{cdf:mt1}. The middle
heptagon is the diagram~\eqref{cd:b2} with $\tr\sm A$ instead of $\tr$
and $\wti S$ instead of $S$; where we again use the identity $\tr\sm
A\dar\wti S=\tr\dar S\sm A$. Since $|\supp(\tr\sm
A)|=|\supp\tr|-|A|<|\supp\tr|$, the induction hypothesis implies that
this heptagon commutes. The leftmost pentagon is \eqref{cd:xs}, with
$\wti S$ instead of $S$, whereas the rightmost pentagon is
\eqref{cd:xs} as well, this time with $\wti S$ instead of $S$, and
$\chi(\tr)$ instead of~$\tr$. They both commute by
Proposition~\ref{prop:cxs}. Again, removing the $2$ inner terms from
the diagram on Figure~\ref{cdf:mt1} will yield the diagram~\eqref{cd:m1} with
$A=T=B$, so we conclude that \eqref{cd:m1} commutes in this special
case.

In general, when $T\subseteq A$, we have a sequence of inclusions
$X_{S,A}\hra X_{S,T}\hra X_{T,T}\hra X_{T,B}$. Again, it is easy to
see that the commutativity of the diagram \eqref{cd:m1} for the
inclusion $X_{S,A}\hra X_{T,B}$ follows from the commutativity of the
diagrams \eqref{cd:m1} for the inclusions $X_{S,A}\hra X_{S,T}$,
$X_{S,T}\hra X_{T,T}$, and $X_{T,T}\hra X_{T,B}$. Hence we are done
with the proof of this case as well.


We now know that $\Phi(\tr)$ is a well-defined homeomorphism between
$P(\tr)$ and $P(\chi(\tr))$. To finish the proof of the main theorem,
we need to check the commutativity of the diagrams~\eqref{cd:b}
and~\eqref{cd:b2}. The commutativity of~\eqref{cd:b2} is an~immediate
consequence of~\eqref{cd:phi}, and the way $\Phi(\tr)$ was defined. To
show that~\eqref{cd:b} commutes, pick any $S\subseteq\act\tr$, which
is disjoint from $A$, and consider the diagram on Figure~\ref{cd:mt2}.
\begin{figure}[hbt]
\begin{tikzcd}
P(\chi(\tr))\arrow[hookleftarrow]{rr}{\alpha}\arrow[hookleftarrow]{dd}{\beta} & 
&X_S(\chi(\tr))\arrow{rr}{\rho}[swap]{\cong} 
& &P(\chi(\tr_S))\arrow[hookleftarrow]{dd}{\beta} \\
&P(\tr)\arrow[hookleftarrow]{d}{\beta}\arrow{ul}{\cong}[swap]{\Phi}\arrow[hookleftarrow]{r}{\alpha}
&X_S(\tr)\arrow[squiggly]{r} \arrow[hookleftarrow]{d}
&P(\tr_S)\arrow{ur}{\Phi}[swap]{\cong}\arrow[hookleftarrow]{d}{\beta} \\
B_V(\chi(\tr))\arrow[squiggly]{dd}{\delta} 
&B_V(\tr)\arrow[squiggly]{d}{\delta}\arrow[hookleftarrow]{r}
&X_{S,\es,V}(\tr)\arrow[squiggly]{d}{\psi}\arrow[squiggly]{r}{\varphi} 
&B_V(\tr_S)\arrow[squiggly]{d}{\delta} 
&B_V(\chi(\tr_S))\arrow[squiggly]{dd}{\delta}\\
&P(\tr\sm V)\arrow{dl}{\cong}[swap]{\Phi}\arrow[hookleftarrow]{r}{\alpha}
&X_S(\tr\sm V)\arrow[squiggly]{r} 
&P(\bar r_{S,V})\arrow{dr}{\Phi}[swap]{\cong} \\
P(\chi(\tr\sm V))\arrow[hookleftarrow]{rr}{\alpha}& 
&X_S(\chi(\tr\sm V))\arrow{rr}{\nu}[swap]{\cong} & 
&P(\chi(\bar r_{S,V}))
\end{tikzcd}
\caption{Commuting diagram used in the proof of Theorem~\ref{thm:main}}
\label{cd:mt2}
\end{figure}
The maps $\varphi$ and $\psi$ are as in Proposition~\ref{prop:6.13},
and the maps $\rho$ and $\nu$ are given by
\[
\begin{tikzcd}[column sep=2cm]
\rho:X_S(\chi(\tr))\arrow[squiggly]{r}{\gamma_S(\chi(\tr))}
&P(\chi(\tr)_S)\arrow{r}{\tau(\chi(\tr_S),\chi(\tr)_S)^{-1}}
&P(\chi(\tr_S))
\end{tikzcd}
\] 
and
\[
\begin{tikzcd}[column sep=2cm]
\nu:X_S(\chi(\tr\sm A))\arrow[squiggly]{r}{\gamma_S(\chi(\tr\sm A))}
&P(\chi(\tr)_{S,A})\arrow{r}{\tau(\chi(\tr_{S,A}),\chi(\tr)_{S,A})^{-1}}
&P(\chi(\tr_{S,A})),
\end{tikzcd}
\] 
where we use the identities $\chi(\tr\sm A)=\chi(\tr)\sm A$,
$\chi(\tr_S)\sm A=\chi(\tr_{S,A})$, and $\chi(\tr\sm A)_S=\chi(\tr)_{S,A}$,
with the latter one relying on the fact that $S\cup A=\es$.

Let us investigate the diagram on Figure~\ref{cd:mt2} in some detail. The
middle part is precisely the diagram~\eqref{eq:bar}, which commutes by
Proposition~\ref{prop:6.13}. We have $4$ hexagons surround that middle
part.  The hexagon on the left is the diagram~\eqref{cd:b} itself. The
hexagon above is precisely the diagram~\eqref{cd:b2}, so it
commutes. The hexagon below is the diagram~\eqref{cd:b2} with $\tr\sm
A$ instead of $\tr$, where we use \eqref{eq:rsa2} again. This diagram
commutes by the induction hypothesis. The hexagon on the right is the
diagram~\eqref{cd:b} with $\tr_S$ instead of $\tr$. Since
$|\tr_S|<|\tr|$, it also commutes by the induction assumption.

Let us now show that the diagram obtained from the one on Figure~\ref{cd:mt2} 
by the removal of the $9$ inner terms commutes. This diagram can be factorized 
as shown on Figure~\ref{cdf:mt3}.
\begin{figure}
\begin{tikzcd}
P(\chi(\tr))\arrow[hookleftarrow]{r}{\alpha}\arrow[hookleftarrow]{d}{\beta} 
&X_S(\chi(\tr))\arrow[squiggly]{r}{\gamma} 
&P(\chi(\tr)_S)\arrow[leftarrow]{r}{\tau}[swap]{\cong}\arrow[hookleftarrow]{d}{\beta} 
&P(\chi(\tr_S))\arrow[hookleftarrow]{d}{\beta} \\
B_V(\chi(\tr))\arrow[squiggly]{d}{\delta} & 
&B_V(\chi(\tr)_s)\arrow[squiggly]{d}{\delta} 
&B_V(\chi(\tr_S))\arrow[squiggly]{d}{\delta}\\
P(\chi(\tr\sm V))\arrow[hookleftarrow]{r}{\alpha} 
&X_S(\chi(\tr\sm V))\arrow[squiggly]{r}{\gamma} 
&P(\chi(\tr)_{S,V})\arrow[leftarrow]{r}{\tau}[swap]{\cong}
&P(\chi(\tr_{S,V}))
\end{tikzcd}
\caption{Commuting diagram used in the proof of Theorem~\ref{thm:main}}
\label{cdf:mt3}
\end{figure}
The left part of the diagram on Figure~\ref{cdf:mt3} is \eqref{eq:bar} with
$\chi(\tr)$ instead of $\tr$, whereas the right part of the diagram on 
Figure~\ref{cdf:mt3} is the diagram~\eqref{cd:tau} with $\chi_{C_1,D_1}=\chi(\tr_S)$,
$\chi_{C_2,D_2}=\chi(\tr)_S$. They both commute, hence so does the
whole diagram.

Consider now two sequences of maps in the diagram on Figure~\ref{cd:mt2}:
\begin{equation}
\label{eq:s1}
\begin{tikzcd}[column sep=0.6cm]
B_V(\tr)\cap X_S(\tr)\arrow[hookrightarrow]{r}
&B_V(\tr)\arrow[hookrightarrow]{r}{\beta}
&P(\tr)\arrow{r}{\Phi}[swap]{\cong}&P(\chi(\tr))
\end{tikzcd}
\end{equation}
and
\begin{equation}
\label{eq:s2}
\begin{tikzcd}[column sep=0.5cm]
X_{S,\es,V}(\tr)\arrow[hookrightarrow]{r}
&B_V(\tr)\arrow[squiggly]{r}{\delta}
&P(\tr\sm V)\arrow{r}{\Phi}[swap]{\cong}
&P(\chi(\tr\sm V))\arrow[leftsquigarrow]{r}{\delta}
&B_V(\chi(\tr))\arrow[hookrightarrow]{r}{\beta}
&P(\chi(\tr))
\end{tikzcd}
\end{equation}
It follows by a~simple diagram chase that the commutativities in the
diagram on Figure~\ref{cd:mt2} which we have shown imply the equality
of these two maps. This is true for all $S$, such that
$S\subseteq\act\tr$ and $S\cap V=\es$. On the other hand, the
subcomplexes $X_{S,\es,V}(\tr)$, where $S\subseteq\act\tr$, $S\cap
V=\es$, cover $B_V(\tr)$.  As a~matter of fact, the simplicial
isomorphisms $\psi$ and $\delta_V(\tr)$ show that they induce
a~stratification which is isomorphic to the stratification of
$P(\tr\sm V)$ by $X_S(\tr\sm V)$. The fact that they cover $B_V(\tr)$
completely implies that the maps \eqref{eq:s1} and \eqref{eq:s2}
remain the same after the first term is skipped, which is the same as
to say that \eqref{cd:b} commutes.  This concludes the proof.  \qed

\begin{crl}\label{crl:main}
For an arbitrary round counter $\tr$ the immediate snapshot complex
$P(\tr)$  is homeomorphic to the closed ball of dimension~$|\supp\tr|-1$.
\end{crl}

\end{document}